\documentclass[10pt,aps,prb,floatfix,superscriptaddress,footinbib,twocolumn]{revtex4-2}

\usepackage[dvipsnames]{xcolor}
\usepackage{hyperref}

\definecolor{webblue}{rgb}{0, 0, 0.5}
\hypersetup{
bookmarks=true,
bookmarksnumbered=false,
bookmarksopen=false,
colorlinks=true,
linkcolor=webblue,
urlcolor=webblue,
citecolor=webblue,
}

\usepackage{bm}
\usepackage{amsmath}
\usepackage{amssymb}
\usepackage{mathtools}
\usepackage{braket}
\usepackage{microtype}
\usepackage{graphicx}
\usepackage{comment}
\usepackage[capitalize]{cleveref}

\newcommand{\vdag}{{\vphantom{\dag}}}
\newcommand{\cre}[1]{{\hat{#1}^\dag}}
\newcommand{\ann}[1]{{\hat{#1}^\vdag}}

\newcommand{\Tc}{T_\mathrm{C}}
\newcommand{\Kp}{{K^\prime}}

\newcommand{\bvec}[1]{\bm{#1}}
\newcommand{\prlparagraph}[1]{{\itshape{#1}}---}
\newcommand{\HOVHS}{\textls[-50]{HOVHS}}
\newcommand{\VHS}{\textls[-50]{VHS}}

\newcommand{\supplement}[1]{%
  \clearpage%
  \title{#1}%
  \maketitle%
  \setcounter{equation}{0}%
  \setcounter{figure}{0}%
  \setcounter{table}{0}%
  \setcounter{page}{1}%
  \makeatletter%
  \renewcommand{\thesection}{S\arabic{section}}%
  \renewcommand{\thesubsection}{\Alph{subsection}}%
  \renewcommand{\theequation}{S\arabic{equation}}%
  \renewcommand{\thefigure}{S\arabic{figure}}%
  \renewcommand{\thetable}{S\Roman{table}}%
  \renewcommand{\thepage}{S\arabic{page}}%
  \renewcommand{\theequation}{SM \arabic{equation}}
  \renewcommand{\thefigure}{SM \arabic{figure}}
  \renewcommand{\thesection}{SM \arabic{section}}
  \makeatother%
  \onecolumngrid%
}

\makeatletter
\def\maketitle{
\@author@finish
\title@column\titleblock@produce
\suppressfloats[t]}
\makeatother

\newcommand{\lk}[1]{{\color{purple} [LK: #1]}}

\renewcommand{\lk}[1]{{\color{red}#1}}

\renewcommand{\autoref}[1]{\cref{#1}}

\begin{document}

\title{Kekul\'e order from diffuse nesting near higher-order Van Hove points}

\author{Jonas Beck}
\thanks{These authors contributed equally.}
\affiliation{Institut für Theoretische Physik und Astrophysik and Würzburg-Dresden Cluster of Excellence ct.qmat, Universität Würzburg, 97074 Würzburg, Germany}
\author{Jonathan Bodky}
\thanks{These authors contributed equally.}
\affiliation{Institut für Theoretische Physik und Astrophysik and Würzburg-Dresden Cluster of Excellence ct.qmat, Universität Würzburg, 97074 Würzburg, Germany}
\author{Matteo Dürrnagel}
\email{matteo.duerrnagel@uni-wuerzburg.de}
\affiliation{Institut für Theoretische Physik und Astrophysik and Würzburg-Dresden Cluster of Excellence ct.qmat, Universität Würzburg, 97074 Würzburg, Germany}
\affiliation{Institute for Theoretical Physics, ETH Z\"urich, 8093 Z\"urich, Switzerland}
\author{Ronny Thomale}
\email{ronny.thomale@uni-wuerzburg.de}
\affiliation{Institut für Theoretische Physik und Astrophysik and Würzburg-Dresden Cluster of Excellence ct.qmat, Universität Würzburg, 97074 Würzburg, Germany}
\author{Julian Ingham}
\email{ji2322@columbia.edu}
\affiliation{Department of Physics, Columbia University, New York, NY 10027, USA}
\author{Lennart Klebl}
%\email{lennart.klebl@uni-wuerzburg.de}
\affiliation{Institut für Theoretische Physik und Astrophysik and Würzburg-Dresden Cluster of Excellence ct.qmat, Universität Würzburg, 97074 Würzburg, Germany}
\author{Hendrik Hohmann}
\affiliation{Institut für Theoretische Physik und Astrophysik and Würzburg-Dresden Cluster of Excellence ct.qmat, Universität Würzburg, 97074 Würzburg, Germany}

\date{\today}

\begin{abstract}

Translation symmetry-breaking order is assumed to be suppressed by the lack of Fermi surface nesting near certain higher-order Van Hove singularities (\HOVHS{}). We show the anisotropic band-flattening inherent to such \HOVHS{}, combined with broadening of the Fermi surface due to elevated critical scales, results in the Fermi surface becoming approximately nested at a wavevector unrelated to the precise shape of the Fermi surface---leading to a $\sqrt{3}\times\sqrt{3}$  Kekul\'e density wave formation. The effect is demonstrated using unbiased renormalization group calculations for a model of the breathing kagome lattice. Our mechanism---termed 
\textit{diffuse nesting}---represents an entirely new notion in the study of Fermi surface instabilities.
\end{abstract}

\maketitle

\prlparagraph{Introduction.}%
When distinct features in the electronic dispersion $\varepsilon(\bvec{k})$ are absent, superconductivity is the unique weak-coupling instability of the Fermi liquid~\cite{shankar1994renormalization}. Notable exceptions occur near Van Hove singularities (VHS)—where the density of states diverges—or when the Fermi surface exhibits the so-called nesting property $\varepsilon(\bvec k) = -\varepsilon(\bvec k+\bvec q)$, in which case spin and charge density wave orders with finite wavevectors $\bvec q$ can emerge and compete with unconventional superconductivity~\cite{Schulz1987, Furukawa1998, Dzyaloshinskii1987, Kiesel2012b, Kiesel2013, Nandkishore2012, scammell2023chiral, ingham2024theory, ingham2025vestigial,fischer2024theory, chubukov2008magnetism, vorontsov2010superconductivity, maiti2013superconductivity}.

The emergence of these ordered states can be understood in terms of the low-temperature behavior of the particle-particle and particle-hole susceptibilities, $\chi_{pp}$ and $\chi_{ph}$, which exhibit logarithmic divergences as the system approaches the critical temperature.
Recently, the circumstances under which unconventional Fermi surface instabilities may arise have received renewed attention with the discovery of various quasi-two-dimensional materials exhibiting unconventional charge order and superconductivity, including untwisted~\cite{jindal2023coupled} and twisted transition metal dichalcogenides (TMDs)~\cite{wang2020correlated, xu2020correlated,  regan2020mott, li2021quantum, huang2021correlated, anderson2023programming, cai2023signatures, xu2023observation, park2023observation, wang2025hidden, ghiotto2024stoner, guo2025superconductivitya, xia2024unconventional}, rhombohedrally stacked~\cite{zhou2021superconductivity, zhou2021half, seiler2022quantum, zhou2022isospin, zhang2023enhanced, tsui2024direct, lu2024fractional} and twisted graphene multilayers~\cite{cao2018correlated,
cao2018unconventional, lu2019superconductors, yankowitz2019tuning, sharpe2019emergent, nuckolls2020strongly, serlin2020intrinsic, shen2020correlated, chen2021electrically, hao2021electric, park2021tunable, saito2021hofstadter, wu2021chern, xie2021fractional, rubio-verdu2022moire}, and kagome metals~\cite{
ortiz2020csv3sb5, yang2020giant, chen2021roton, jiang2021unconventional, li2021observation, zhao2021cascade, hu2022rich, hu2022topological, khasanov2022timereversal, mielke2022timereversal, nie2022chargedensitywavedriven, hossain2025field, teng2022discovery, xu2022threestate, jiang2023flat, li2023electronic, yang2023observation, yang2024superconductivitya, guo2024correlated, bigi2024pomeranchuk, nag2024pomeranchuk, jiang2024van, kang2022twofold}.
Experimental and theoretical studies have shown that some of these systems host a variety of \VHS{} known as a higher-order Van Hove singularity (\HOVHS{})~\cite{irkhin2002robustness, McChesney2010, yudin2014fermi, Shtyk2017, NatComYuan, Chandrasekaran2020, kang2022twofold, Chandrasekaran2023a, Chandrasekaran2023b, wang2025higher, guerci2024topological, perkins2025designing, classen2024high, classen2020competing}, where the dispersion flattens significantly. In contrast to the logarithmic divergence characteristic of a conventional \VHS{}, this leads to an algebraically diverging density of states (DOS). 

From a technical perspective, the transition from a logarithmic to an algebraic divergence of the DOS substantially alters the weak coupling picture for \HOVHS, as $\chi_{pp}$ and $\chi_{ph}$ likewise attain algebraic divergences. For many such saddle points, the anisotropy of the band flattening suppresses the nesting condition, rendering the Fermi liquid unstable to translationally invariant particle-hole instabilities at \textit{zero} momentum transfer, rather than density wave order, i.e. $\bvec q = 0$ Pomeranchuk instabilities such as nematic order or altermagnetism~\cite{classen2020competing, Han2023, nag2024pomeranchuk, patra2025high}. %The study of such phases has enriched our understanding of the possible patterns of competing orders in metallic systems~\cite{classen2024high}.

In this Letter, we propose an unprecedented mechanism for the formation of translation-symmetry breaking $\sqrt3\times\sqrt3$ density waves in hexagonal lattice systems
with \HOVHS{}. The phenomenon arises from the interplay between warped Fermi surfaces induced by anisotropic \HOVHS{} and elevated critical scales, leading to what we term \textit{diffuse nesting}. That is, the associated nesting condition manifests only when the Fermi surface retains a finite broadening, e.g. due to finite temperature, interactions, or potentially disorder. From an itinerant perspective, a high density of states implies elevated critical scales of ordered states, which makes this condition natural in systems with \HOVHS{}. Remarkably, the resulting instabilities are characterized by a wavevector $\bvec{q} = K$ unrelated to the momenta connecting the \HOVHS{} at the $M$-point, and so do not involve the regions of the Fermi surface where the density of states is largest.

The resulting state is a variant of Kekulé order, first conceived as the ground state of Benzene. It describes an alternating hexagonal $\sqrt{3}\times\sqrt{3}$ bond order,
previously discussed in the context of carbon nanotubes, graphene~\cite{mintmire1992fullerene, chamon2000solitons, hou2007electron, nomura2009field, cheianov2009ordered, weeks2010interaction, gomes2012designer}, and more recently twisted graphene multilayers~\cite{thomson2018triangular, kwan2021kekule, wagner2022global, wang2023groundstate, ingham2023quadratic, kwan2024electron, wang2024kekule}. The similarity to the Kekul\'e orders previously studied in honeycomb systems is emphasized by examining the line graph reconstruction of the kagome lattice, which reveals a Kekul\'e-Y pattern on the resulting effective honeycomb structure.

Our findings are illustrated by functional renormalization group calculations~\cite{Salmhofer2001, Metzner2012, Platt2013f, Profe2022, Beyer2022, Profe2024div1, Profe2024div2} for a model inspired by the breathing kagome surface of ferromagnetic Co\textsubscript{3}Sn\textsubscript{2}S\textsubscript{2}, which features \HOVHS{} near the Fermi level \cite{nag2024pomeranchuk}. We trace the origin of the $K$-ordering tendency to specific properties inherent to such \HOVHS{} and propose a simplified single-band model, featuring an eigenspectrum solely determined by series-expanded saddle points situated at the $M$-points---thus generalizing our observation to arbitrary hexagonal lattice systems. While we focus on a model of spinless fermions---in which any density wave necessarily occurs in the charge channel---our results can be extended to spinful systems, revealing magnetic Kekul\'e-type order in the breathing kagome configuration, as detailed in the Supplemental Material (SM) \cite{SM}.

Many prior studies of higher-order Van Hove points have worked with so-called $g$-ology or patch models, which approximate the Fermi surface as isolated points at the \VHS{}~\cite{classen2020competing, Han2023, castro2023emergence, YiMingWu_2023, hsu2021spin}. Our results dramatically illustrate that such models necessarily fail to capture certain ordering tendencies of a \HOVHS{}, which counter-intuitively involve the fermions \emph{away} from the momenta where the density of states is largest, reminiscent of the imperfect nesting scenario in Ref.~\cite{Johannes2008}. The phenomenon we describe represents an entirely new notion in the study of Fermi surface instabilities.

\begin{figure}%
    \centering
    \includegraphics[width=1\linewidth]{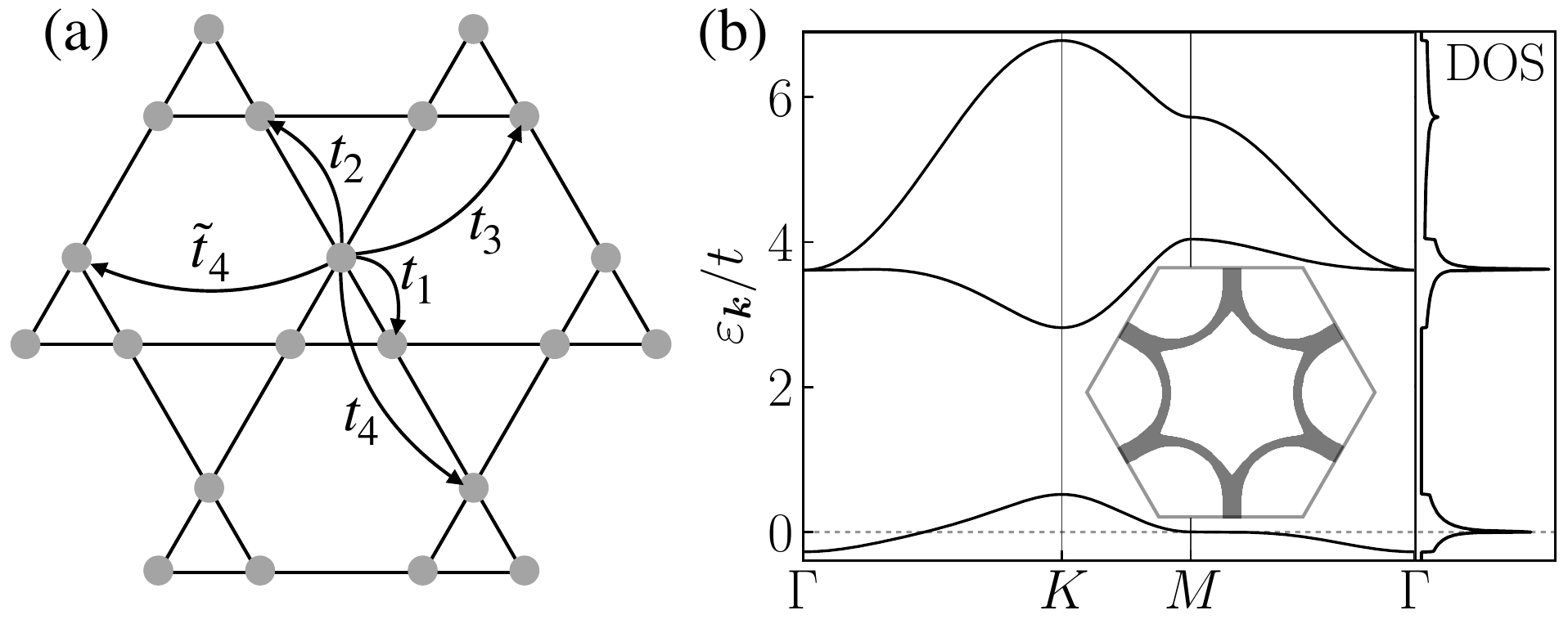}
    \caption{Breathing kagome model. (a)~Breathing kagome lattice configuration with hoppings up to fourth nearest neighbor. (b)~Corresponding band structure with density of states, characterized by an algebraic divergence at the Fermi level. The inset displays the broadened Fermi surface at $T=0.01\,t$.
    }
    \label{fig:1}
\end{figure}

\prlparagraph{Model.}%
\HOVHS{}  can emerge in tight-binding models when longer-ranged or complex hoppings are introduced. In parameter regimes where the quadratic term of the dispersion near the Van Hove point is suppressed, a quartic dispersion emerges~\cite{classen2020competing, Han2023}. One concrete realization is the breathing configuration of the kagome Co\textsubscript{3}Sn surface termination (\cref{fig:1}a) of ferromagnetic shandite compound Co\textsubscript{3}Sn\textsubscript{2}S\textsubscript{2}~\cite{nag2024pomeranchuk}. The tight-binding model adapted from \textit{ab initio} calculations reads 
\begin{multline}
    \mathcal{H} = -\mu \sum_{ i} n_{i} \, + \, \sum_{n=1}^4 t_n \sum_{\langle i,j\rangle_n} \cre{c}_{i} \ann{c}_{j}
    +\sum_{n=1}^4 \,V_n \sum_{\langle i,j\rangle_n} n_{i} n_{j} \,  ,
\label{eqn:Hamiltonian}
\end{multline}
where $c_{i}$ annihilates an electron at site $i$, $n_{i}=c_{i}^{\dagger}\,c^{\phantom{\dagger}}_{i}$ is the electron density operator and $\mu$ the chemical potential; $t_{1\rightarrow 4} \approx (-1.383, -0.617, 0.352, -0.176)\,t$ correspond to a choice of hopping amplitudes up to fourth nearest neighbor with energy scale $t$, producing an exact \HOVHS{} (\cref{fig:1}a). These parameters deviate slightly from the fitted \textit{ab initio} calculation, see \autoref{sec:kinetic_models_breathing_kagome} and \autoref{sec:stability_analysis}~\cite{SM} for details and stability analyses. The hopping $t_4$ vanishes along a particular direction for each sublattice, indicated as $\tilde{t}_4$ in \cref{fig:1}a. In addition to the kinetic term, we add repulsive density-density interactions $V_n$.

The resulting dispersion, which features a \HOVHS{}  at the $M$-points, is shown in \cref{fig:1}b, along with the density of states (DOS) hosting an algebraic divergence at the Fermi level; a notable difference can be observed between the DOS of the \HOVHS{} and that of the ordinary \VHS{} near $\varepsilon \approx 6t$ in the upper band. An additional \HOVHS{} emerges at $\bvec{q}=\Gamma$, where the two upper bands touch. 
The inset displays the Fermi surface (FS) for a finite window of $\mu$. Due to the anisotropic higher-order nature of the dispersion along the $\Gamma M$ direction---which becomes proportional to $ \alpha \, k^2_{K M} - \beta\, k^4_{\Gamma M}$, where $k_{ab}$ denotes the momentum component along direction $ab$---the FS becomes significantly flattened. As a consequence, the perfect nesting at transfer momentum $M_i$, which is typical for ordinary \VHS{} in hexagonal lattice systems, is lifted. Although the majority of the DOS remains concentrated at the $M$-points, the band flattening induces an enhanced spread along the Fermi surface (FS) surrounding $M$. 

\begin{figure}
    \centering
    \includegraphics[width=1\linewidth]{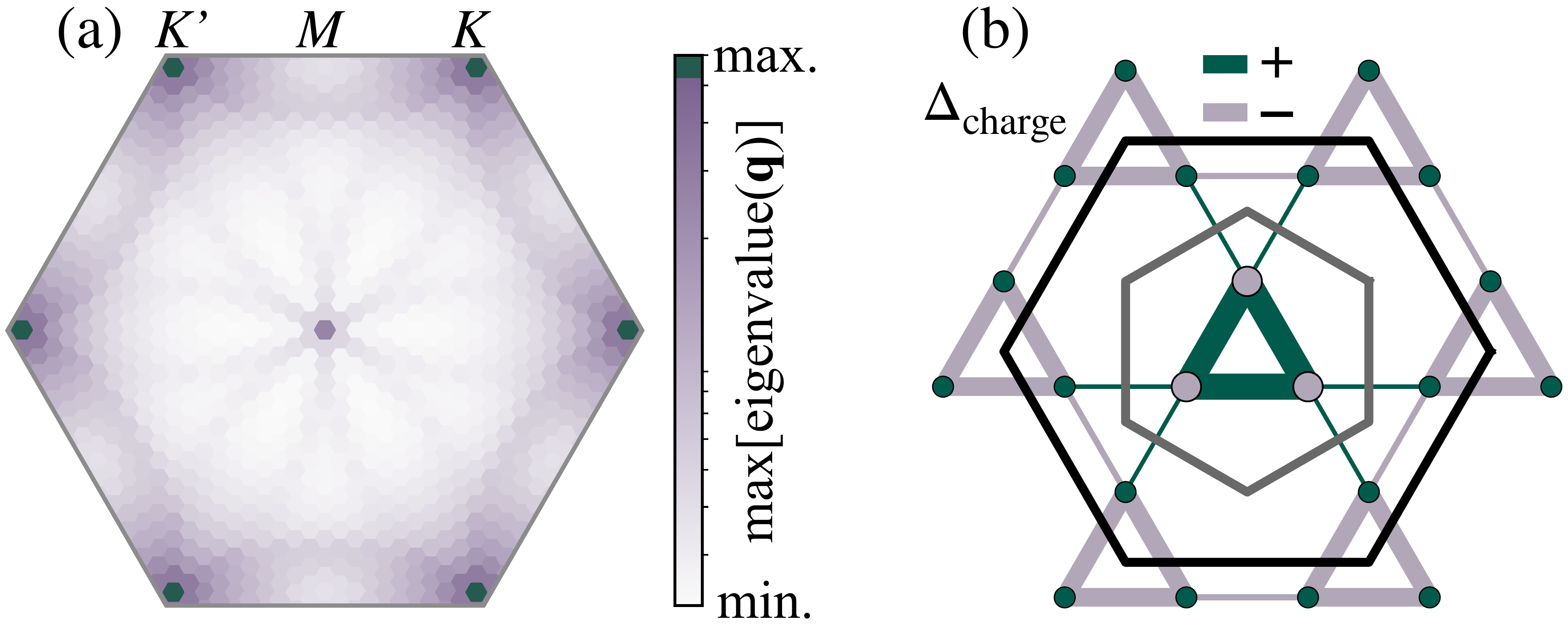}
    \caption{Leading eigenvalues and order parameter.
    (a)~Leading eigenvalues of the effective scattering vertex in the particle-hole channel on a transfer momentum $\boldsymbol{q}$-mesh in the Brillouin zone. The dominant eigenvalues are situated at the two inequivalent $K$-points. (b)~Order parameter obtained by combining functional renormalization group calculations with Landau-Ginzburg and mean-field analysis.
    }
    \label{fig:evals_k}
\end{figure}

\prlparagraph{Kekul\'e order.}%
To analyze FS instabilities of the breathing kagome model of spinless fermions, we employ the functional renormalization group (FRG), which is an unbiased method for the treatment of competing orders in the particle-particle and particle-hole channels~\cite{Salmhofer2001, Metzner2012, Platt2013f, Profe2022, Beyer2022, Profe2024div1, Profe2024div2}.
FRG successively integrates out interaction processes at energies above a cutoff scale $\Lambda$, which is lowered from a high initial value down to zero. This process gradually captures low-energy physics, thereby naturally probing the FS at various degrees of broadening~\cite{fischer2024theory}. Formally, the emergence of an ordered phase is indicated by a divergent eigenvalue of the effective two-particle interaction vertex at a critical scale $\Lambda_c$, with the corresponding eigenstate defining the structure of the order parameter (see SM \cite{SM} for further details on FRG and mean-field decomposition of the four-point interaction vertex). The critical scale $\Lambda_c$ serves as an indicator of the critical temperature, provided that an appropriate regulator is used (e.g. a sharp frequency cutoff \cite{Metzner2012, Profe2024}). Within the FRG framework employed, the (bosonic) transfer momenta $\boldsymbol{q}$ are parametrized on a high-resolution momentum grid that covers the entire Brillouin zone (BZ), visualized by the plot of the leading eigenvalue of the two-particle vertex as a function of $\bvec{q}$ in \cref{fig:evals_k}a, shown for $\Lambda$ greater the critical scale at which the eigenvalue diverges.

While the pristine kagome Hubbard model, featuring perfect nesting and sublattice interference at its Van Hove fillings, has been studied extensively using FRG methods~\cite{Kiesel2013, Wang2013, Profe2024}, the influence of \HOVHS{} has mainly been investigated either by mean-field theories \cite{NatComYuan}, or by renormalization group approaches applied to simple models known as patch theories~\cite{classen2020competing, Han2023, hsu2021spin, YiMingWu_2023}. Relying on the assumption that the relevant fermions are those where the density of states is largest, these simple models discard the Fermi surface away from the vicinity of the \HOVHS{}. For \HOVHS{} near $M$, the FS scattering processes are then restricted to transfer momenta $\boldsymbol{q} = (M,\,\Gamma)$. However, by relaxing the constraint to transfer momenta which relate \HOVHS{} points within our FRG analysis, we observe an unprecedented ordering tendency towards
$\bvec q=K$ density waves---signified by the dominant peaks at $K$ in \autoref{fig:evals_k}a---with significant bond character, i.e. Kekul\'e order. Even though this wavevector is unrelated to the position of the \HOVHS{}, we observe the Kekul\'e instability in an extensive parameter regime (see SM~\cite{SM}).

This instability of the particle-hole channel breaks translation symmetry, and corresponds to the two inequivalent wavevectors $K$ and $K'$ related by time-reversal symmetry (\autoref{fig:evals_k}a). The enlarged $\sqrt{3}\times \sqrt{3}$ periodicity triples the size of the unit cell (\cref{fig:evals_k}b, black). The resulting order parameter%
\begin{equation}%
    \Delta^{\bvec q}_{o_1 o_2}(\bvec k) = \frac{1}{N_{\bvec k}} \!\sum_{\bvec k' o_3 o_4}\!
        V^D_{o_1 o_2 o_3 o_4}(\bvec q, \bvec k, \bvec k') \langle \cre{c}_{\bvec k' -\bvec q o_3} \ann{c}_{\bvec k' o_4} \rangle ,
\end{equation}%
is given by the eigenstate of the two-particle vertex $V^D$, written in the direct particle-hole channel ($D$ channel), and corresponds to the $E'_3$ irreducible representation of the extended point group $C''_{3v}$ (see SM~\cite{SM}).

As the symmetries of the initial Hamiltonian are preserved during the FRG flow, our approach is valid only upon transition into the symmetry-broken phase. A subsequent mean-field (MF) analysis is employed to determine the physically realized linear superposition of the two degenerate $E'_3$ basis states $\Delta^{K/K'}$ in the ordered phase~\cite{Reiss2007, Metzner2012, Platt2013f}
\begin{equation}
    \Delta_{\boldsymbol{k}}(A,\phi) = A\,\big[ e^{i \phi}  \Delta_{\boldsymbol{k}}^K + e^{-i \phi} \Delta_{\boldsymbol{k}}^{\Kp} \big] \, ,
\label{main_eqn:order_parameter}
\end{equation}
where the equal amplitude and opposite phase of the two terms are dictated by hermiticity.  Minimizing the MF free energy \cite{SM} reveals minima for $\phi= 0$ and $\phi=\pi$, with $\phi=\pi$ being the global free energy minimum for all amplitudes $A$ for $T<T_c$. The real-space representation of \cref{main_eqn:order_parameter} for $\phi=\pi$ is shown in \cref{fig:evals_k}b. The thickness of the bonds/dots represents the absolute value, with green (purple) indicating a positive (negative) value of the order parameter corresponding to an electron charge accumulation (depletion) within our model of spinless fermions.

It is noteworthy that our kagome Kekul\'e phase exhibits real-space features similar to those previously reported in honeycomb models.
The \VHS{} near the Fermi level in Co\textsubscript{3}Sn\textsubscript{2}S\textsubscript{2} is the remnant of the so-called \textit{mixed} ($m$)-type \VHS{} in the undistorted lattice~\cite{nag2024pomeranchuk}, near which the wavefunction at each $M$-point resides on two of the three kagome sublattices. The local density of states for such a Fermi surface in fact forms a honeycomb lattice, situated at the centers of the corner-sharing triangles~\cite{ingham2025group}. Upon triggering a breathing distortion, the density of states depletes on one sublattice of this emergent honeycomb lattice; the result is that the Fermi level density of states near the \HOVHS{} forms a partially sublattice-imbalanced honeycomb lattice, as can be seen in the nematic density of states simulated in Ref.~\cite{nag2024pomeranchuk}. In real-space, our Kekul\'e state therefore closely resembles the Kekul\'e orders found in honeycomb systems, but with an imbalance between the $A$ and $B$ sublattices. This correspondence can also be visualized through the line graph reconstruction of the kagome lattice, in which the lattice sites can be associated with the bonds of the emergent honeycomb lattice. This construction reveals a Kekul\'e-Y order and implies that the state given in \cref{fig:evals_k}b transforms in the same irreducible representation (see SM~\cite{SM}).

\begin{figure}%
    \centering
    \includegraphics[width=1.0\linewidth]{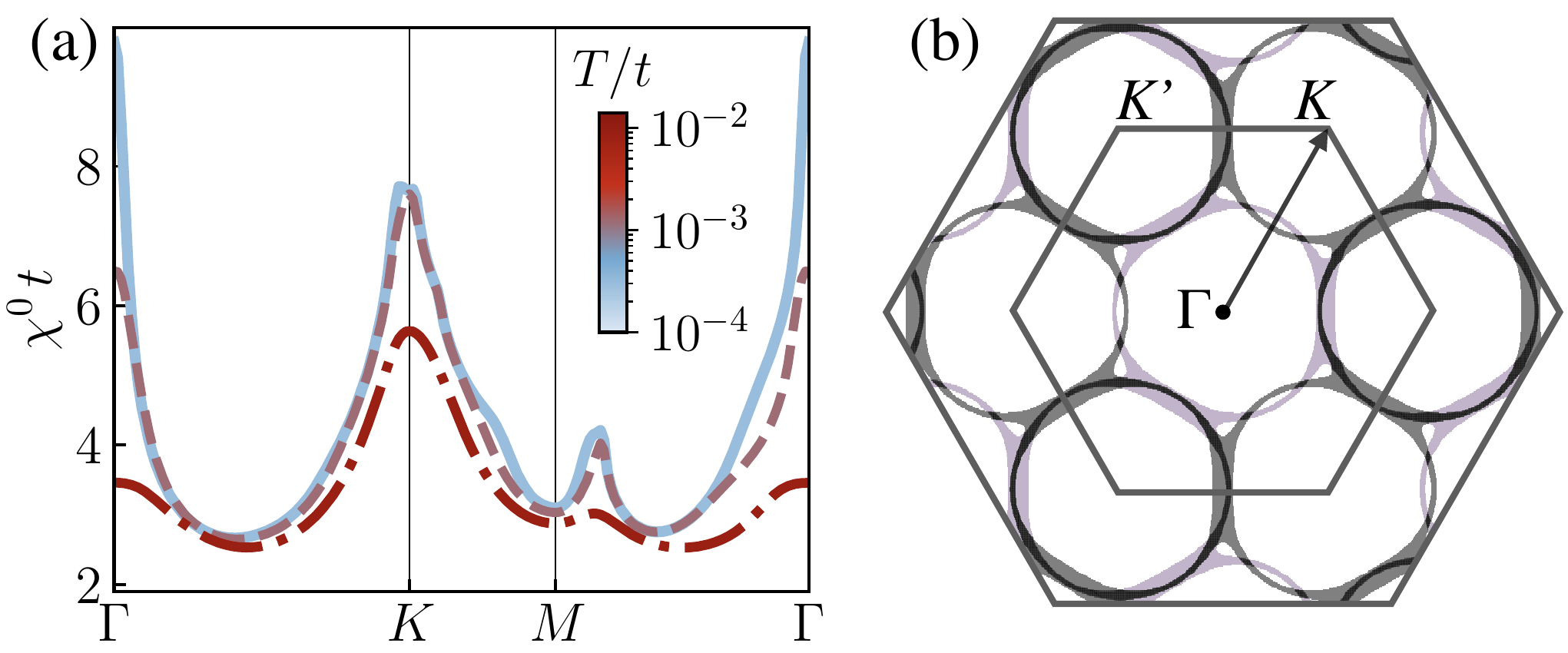}
    \caption{
    Susceptibility and Fermi surface nesting in the kinetic breathing kagome model with FS broadening modeled by temperature. (a)~Maximal eigenvalue of the particle-hole susceptibility \cref{eqn:bare_susc} for three different temperatures $T$. At higher $T$, the peak at momentum $K$ surpasses the $\Gamma$ peak. (b)~Overlap (black) of initial (gray) and $K$-shifted (purple) Fermi surface for $T=10^{-2}\,t$. Circular nesting features drive $K$-order at finite $T$.
    }
    \label{fig:3}
\end{figure}

\prlparagraph{Diffuse nesting.}%
The emergence of $K$-order across a broad interaction parameter regime indicates that the initial structure of the interaction vertex is largely irrelevant to the underlying ordering mechanism, implying that its origin can be discerned from a kinetic viewpoint, i.e. the identification of a
generalized nesting condition.  We consider the static particle-hole susceptibility at finite momentum and temperature \cite{Duerrnagel2022},
\begin{multline} \label{eqn:bare_susc}
    \chi^0_{o_1o_2o_3o_4}(\bvec{q}) = {} \\
    - T\sum_{\omega_n}
    \int_\text{BZ}\,\frac{\text{d}\bvec k}{V_\text{BZ}}
    G_{o_2o_4}(\bvec{k}, \omega_n)
    G_{o_3o_1}(\bvec{k} + \bvec{q}, \omega_n)
    + \text{h.c.}\,,
\end{multline}
where $G_{o_1o_2}(\bvec k, \omega_n)$ is the single particle propagator with momentum $\bvec k$ and fermionic Matsubara frequency $\omega_n$ and $o_i$ indicates the sublattice degree of freedom. 
\Cref{fig:3}a shows that at low temperature, the susceptibility peak at $\bvec{q}=\Gamma$ dominates---in line with the expectation that translationally invariant orders should dominate at weak coupling. Yet at finite FS broadening, which in the kinetic model can be modeled with temperature (see \autoref{sec:nesting}~\cite{SM} for a more detailed discussion of FS broadening), the susceptibility at the transfer momentum $\bvec{q}=K$ exceeds that at $\bvec{q} = \Gamma$.

One can understand Fermi surface nesting as the overlap of the FS with itself upon shifting by a momentum $\bvec{q}$. For an unbroadened FS (zero temperature, no interactions/disorder), no significant overlap can be discerned for nonzero $\boldsymbol{q}$ nesting configurations, owing to the different curvatures of overlapping portions of the FS. 
This changes for a broadened FS, which we realize by introducing a finite temperature $T$ that leads to ``smeared out''{} states near the Fermi level (\autoref{fig:1}b, inset) due to Fermi-Dirac statistics.
For an exemplary broadening of $T=10^{-2}\,t$, \cref{fig:3}b shows nested circular regions (black) that emerge when the broadened FS is shifted by $\boldsymbol{q} = K$, revealing the origin of the dominant finite-momentum peak in the susceptibility displayed in \autoref{fig:3}a. A detailed discussion and visualization for various levels of broadening and nesting vectors is provided in the SM~\cite{SM}.

We emphasize that this argument generalizes beyond our finite temperature calculations, and is expected in the presence of other kinds of Fermi surface broadening, such as that due to interactions or disorder. 
Finite interactions of intermediately correlated systems allow FS scattering processes to access an increasing energy window around the Fermi level comparable to the broadening in \cref{fig:3}b.
As a consequence, we observe $K$-order across the predominant part of the finite interaction parameter regime within our FRG analysis (see SM~\cite{SM}).

Previous studies suggest enhanced nematic ordering tendencies associated with \HOVHS{}~\cite{classen2020competing, Han2023, nag2024pomeranchuk}. These prior models are valid in the strict weak coupling regime, and indeed we observe competing $\boldsymbol{q}=\Gamma$ Pomeranchuck fluctuations---reflected by the peak at $\Gamma$ in \cref{fig:evals_k}a---which become dominant for very weak interaction parameters (FRG results shown in 
\cref{sec:phase_diagrams}~\cite{SM}). Note that the enhanced DOS of \HOVHS{} effectively places the system in an intermediate-coupling regime even for relatively small interactions---further disfavoring the applicability of previous approaches. This renders the functional renormalization group---valid from the weakly to intermediately coupled limits---a vital tool to appropriately capture the leading ordering tendency.

\prlparagraph{Generality.}%
\begin{figure}%
    \centering
    \includegraphics[width=1.0\linewidth]{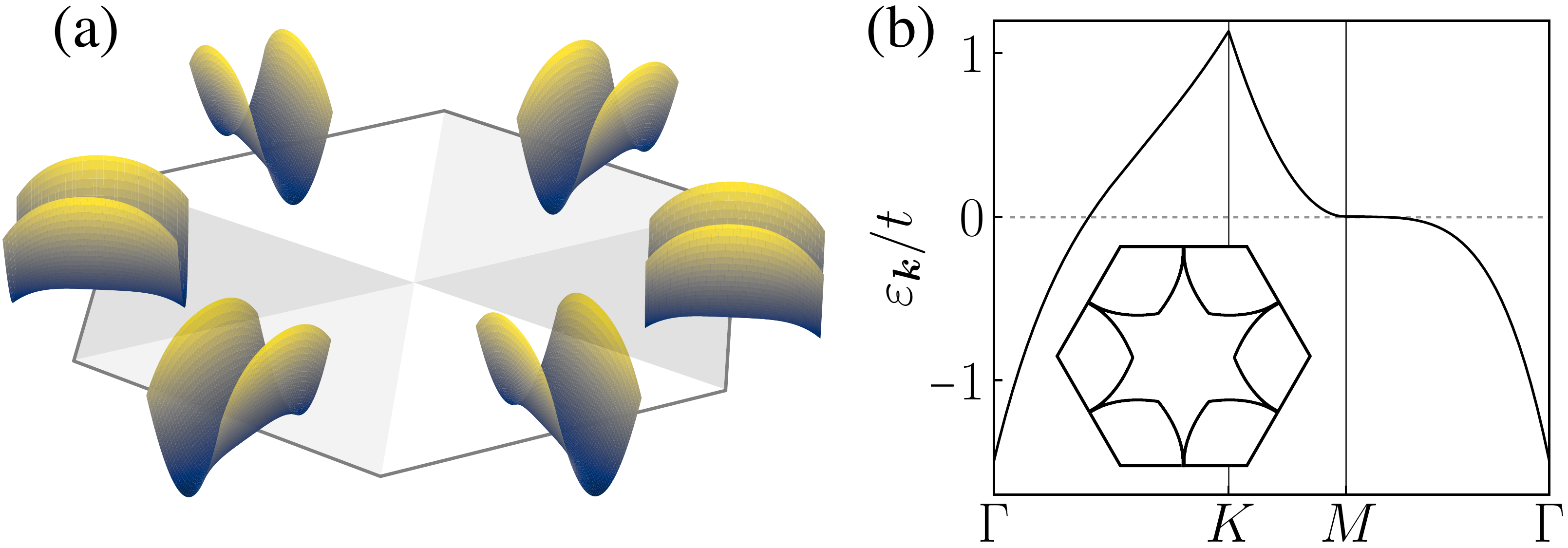}
    \caption{Generalized model. (a)~Taylor expanded higher-order Van Hove dispersion situated at the $M$-points of the hexagonal Brillouin zone. (b)~Corresponding bandstructure and Fermi surface at higher-order Van Hove filling.
    }
    \label{fig:4}
\end{figure}%
The breathing distortion in the model of \cref{eqn:Hamiltonian} effectively introduces a Dirac mass, resulting in a topologically trivial gap opening at $\bvec k=K^{(\prime)}$ (see \cref{fig:1}b).
To explore the generality of the observed Kekul\'e phase and its underlying mechanism, we consider a pristine kagome lattice without breathing distortion that preserves the exact characteristics of the higher-order band flattening. As detailed in the SM~\cite{SM}, we perform FRG calculations for a \textit{mixed} ($m$)-type \HOVHS{} of the kagome lattice. Although $\sqrt{3} \times \sqrt{3}$ ordering remains present in a sizable phase space segment, the extent of this region is reduced. We attribute the suppression to sublattice polarization effects that emerge along the FS due to quantum interference in the pristine kagome lattice~\cite{Kiesel2012a}, as the reintroduction of $\sigma_v$ mirror symmetry restores the so-called sublattice interference effect at the $M$-points. The resulting distinct sublattice occupation along the FS, inherent to the $m$-type~\cite{Kiesel2012a} \HOVHS{}, renders certain momenta inaccessible to local scattering events (see SM~\cite{SM}). The concurrent suppression of critical scales, naturally arising from sublattice polarization effects, further weakens the diffuse nesting mechanism.

An analysis of the quantum geometry associated with the band hosting the \HOVHS{} reveals that, provided the mass term remains sufficiently small, the quantum interference effects between the sublattices remain strong.
In contrast, within the Co\textsubscript{3}Sn\textsubscript{2}S\textsubscript{2} model, the amplitude of the quantum metric $g(\boldsymbol{k})$ is significantly reduced and spread throughout the BZ. Its BZ integral, related to the gauge-independent quadratic spread of Wannier functions 
\begin{equation}
    \Omega_{\mathrm{I}} =\int_{\mathrm{BZ}} \frac{d^2\boldsymbol{k}}{V_{\mathrm{BZ}}} \, \mathrm{Tr}\, [g(\boldsymbol{k})],
\end{equation}
correspondingly decreases, allowing for more localized Wannier functions~\cite{yu2025quantumgeometryquantummaterials, witt2025quantumgeometrylocalmoment} (see \autoref{sec:QGT}~\cite{SM} for details and a quantitative illustration). This allows for a conceptually generalized model consisting of a piecewise series-expanded dispersion around the $M$-points of a hexagonal BZ with quartic dispersion along the $\Gamma M$ direction, as visualized in \cref{fig:4}a. The shaded triangular regions represent the sections of the BZ where the individual expansions govern the dispersion. This piecewise definition preserves the lattice symmetry and ensures a continuous spectrum \cite{SM}.
The resulting bandstructure of the effective single orbital model and its FS are displayed in \cref{fig:4}b. 
Apart from the cusps at the $\Gamma K$ line, the FS retains the same directional band flattening features around the $M$-points. Consequently, diffuse nesting persists, and the dominance of $K$-order is observed within our unbiased FRG analysis, generalizing our proposed diffuse nesting mechanism to arbitrary lattice systems featuring such \HOVHS{}.

\prlparagraph{Conclusion.}%
Our results demonstrate a new mechanism for Fermi surface instabilities: broadening of the Fermi surface introduces approximate nesting, which we term \textit{diffuse nesting}. By tracing the mechanism of $\sqrt{3} \times \sqrt{3}$ density wave formation to the properties of anisotropic band flattening near a \HOVHS{}, we demonstrate the universality of the proposed diffuse nesting phenomenon and consequently expect it to play a crucial role in elucidating the emergent phases in systems with weak to intermediate correlations with \HOVHS{} near the Fermi level. Exemplified by the study of a pristine kagome model featuring a mixed-type \HOVHS{}, we show that sublattice polarization effects can partially suppress the proposed nesting mechanism while preserving a residual tendency towards $\sqrt{3} \times \sqrt{3}$ ordering.

The particular Fermi surface instability we find is reminiscent of Kekul\'e order, where real-space similarities between our kagome Kekul\'e phase and those previously studied in honeycomb models can be discerned. The pronounced enhancement of $K$-nesting implied by the breathing kagome configuration---which can be understood as a deformation of a parent kagome structure as a result of a $C_2$-breaking soft phonon mode---could hint towards cooperative effects between phonon- and Coulomb-driven formation of Kekulé orders in a broader family of correlated systems, including graphene multilayers \cite{nuckolls2023quantum, kim2023imaging, liu2024visualizing} and certain kagome metals \cite{arachchige2022charge, ortiz2024stability, wang2025formation}.

\medskip

\prlparagraph{Acknowledgements.}%
We thank M.~T.~Bunney, A.~Fischer, and D.~Muñoz-Segovia for useful discussions.
M.D. is grateful for support from a Ph.D. scholarship of the Studienstiftung des deutschen Volkes.
This work is supported by the Deutsche Forschungsgemeinschaft (DFG, German Research Foundation) through Project-ID 258499086 -- SFB 1170 and through the W\"urzburg-Dresden Cluster of Excellence on Complexity and Topology in Quantum Matter -- ct.qmat Project-ID 390858490 -- EXC 2147.
We are grateful for HPC resources provided by the Erlangen National High Performance Computing Center (NHR@FAU) of the Friedrich-Alexander-Universit\"at Erlangen-N\"urnberg (FAU). NHR funding is provided by federal and Bavarian state authorities. NHR@FAU hardware is partially funded by the DFG -- 440719683. JI is supported by NSF Career Award No. DMR-2340394.

\let\oldaddcontentsline\addcontentsline
\renewcommand{\addcontentsline}[3]{}

\bibliography{bibliography.bib}

% \clearpage\widetext
\newpage \onecolumngrid

\let\addcontentsline\oldaddcontentsline% Restore \addcontentsline
\setcounter{equation}{0}
\setcounter{table}{0}
\setcounter{section}{0}
\setcounter{figure}{0}
\makeatletter
\renewcommand{\theequation}{SM \arabic{equation}}
\renewcommand{\thefigure}{SM \arabic{figure}}
\renewcommand{\thesection}{SM \arabic{section}}
%\renewcommand{\citenumfont}[1]{S#1}
%\begin{widetext}

\supplement{Supplementary Material: \\
Kekul\'e order from diffuse nesting near higher-order Van Hove points}
\tableofcontents
% Assemble supplement sections
\section{Kinetic models with higher-order Van Hove singularities}

\label{sec:kinetic_models}

\subsection{Breathing kagome model}\label{sec:kinetic_models_breathing_kagome}
The kinetic model we study is inspired by a tight-binding model describing the breathing-distorted kagome surface of Co$_3$Sn$_2$S$_2$ \cite{nag2024pomeranchuk}. The Hamiltonian reads
\begin{equation}
    \mathcal{H} = -\mu \sum_{i, \sigma} \hat{n}_{i,\sigma} \, + \, \sum_{n=1}^4 t_n \sum_{\langle i,j\rangle_{n}, \sigma} \cre{c}_{i,\sigma} \ann{c}_{j,\sigma} \, ,
\label{eqn:Hamiltonian_SM}
\end{equation}
where $\langle i,j\rangle_n$ denotes a summation over $n$-th nearest neighbors with corresponding hopping coefficient $t_n$. The operator
$\cre{c}_{i, \sigma} $ ($\ann{c}_{i, \sigma}$) creates (annihilates) an electron with spin $\sigma$ at site $i$. The electron density operator $\hat{n}_{i, \sigma} = \cre{c}_{i, \sigma} \ann{c}_{i, \sigma}$ at site $i$ with spin $\sigma$ couples to the chemical potential $\mu$. Although the surface of Co$_3$Sn$_2$S$_2$ \cite{nag2024pomeranchuk} is spin polarized, we do not restrict our analysis to the spinless case, where the spin degree of freedom can be regarded as frozen, and the sums over $\sigma$ are omitted. In the sum over the fourth-nearest neighbors, we exclude the hoppings across the hexagons. The corresponding Fourier-transformed Hamiltonian in the spinor basis of the three kagome sublattice sites is given by
\begin{gather}
\mathcal{H}(\bvec{k}) = 
\begin{pmatrix}
2t_4 \text{Re} \gamma_2 +2t_4 \text{Re} \gamma_3 & t_1 + t_2 \bar{\gamma_2} + t_3 \gamma_1 + t_3 \bar{\gamma_3}   & t_1 + t_2 \bar{\gamma_3} + t_3 \bar{\gamma_1} + t_3 \bar{\gamma_2}  \\ 
*& 2t_4 \text{Re} \gamma_1 + 2t_4 \text{Re} \gamma_2 & t_1 + t_2 \bar{\gamma_1} + t_3 \gamma_2 + t_3 \bar{\gamma_3} \\ 
*& *& 2t_4 \text{Re} \gamma_1 + 2t_4 \text{Re} \gamma_3
\end{pmatrix}
\end{gather}
where the hopping phases are $\gamma_1 = \exp(i\bvec{k}\cdot \bvec{R}_1)$, $\gamma_2 = \exp(i\bvec{k}\cdot \bvec{R}_2)$, and $\gamma_3 = \exp(i\bvec{k}\cdot (\bvec{R}_1 + \bvec{R}_2))$, and the lattice vectors are defined as $\bvec{R}_1 = a(1,0)$, $\bvec{R}_2=a(-\tfrac{1}{2},\tfrac{\sqrt{3}}{2})$ with $a$ being the lattice constant. Entries indicated by an asterisk are fixed by hermiticity.

\begin{figure}[b!]
\begin{center}
\includegraphics[width=0.52\columnwidth]{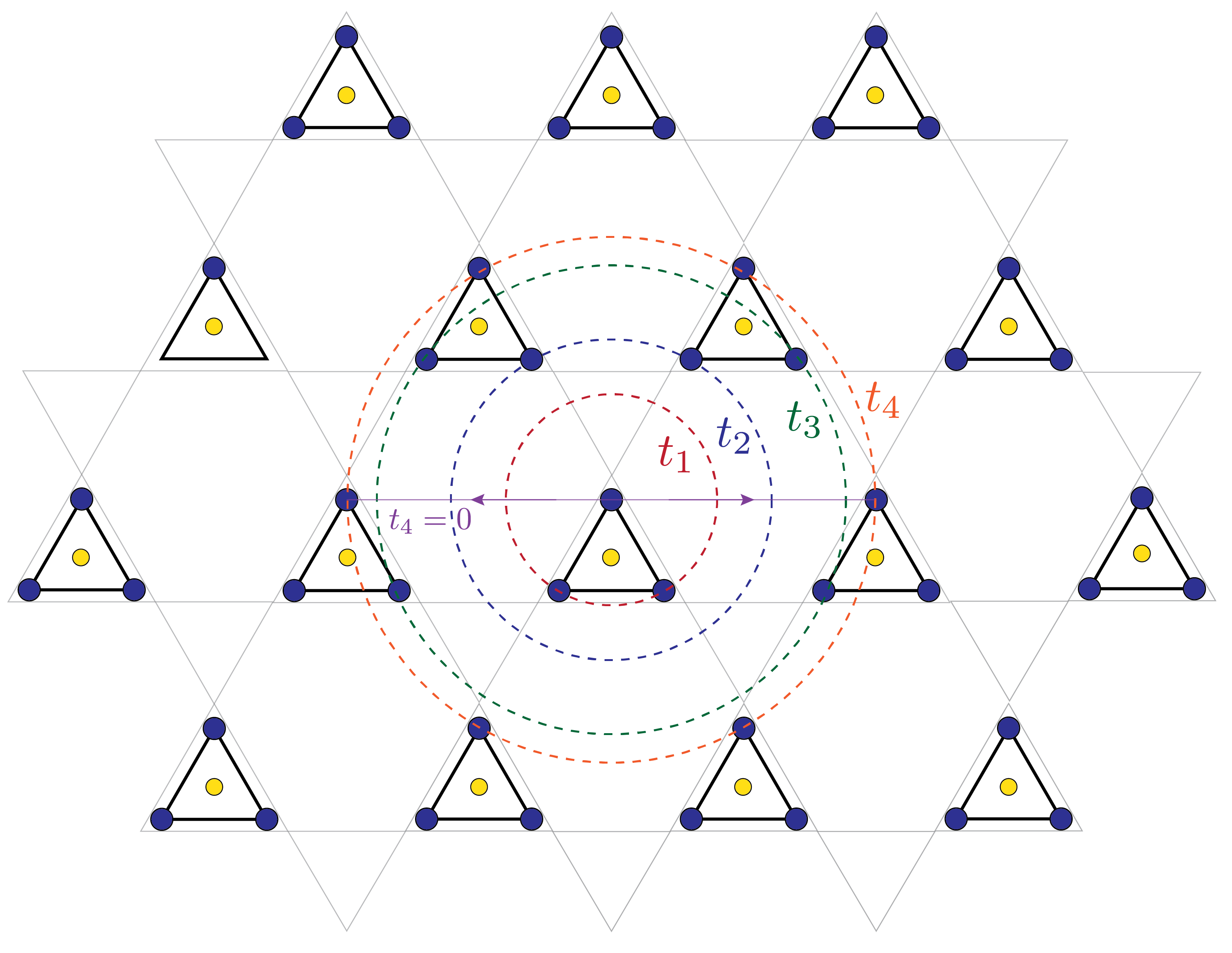}
\end{center}
\vspace{-0.3cm}
\caption[]{Kinetic model for the breathing kagome lattice. The hopping processes $t_{1,2,3,4}$ from sublattice $A$ to the neighbouring unit cells are shown with circles illustrating the hopping distance. The hopping $t_4$ vanishes along a line perpendicular to the residual mirror symmetry at each sublattice (purple). In Co$_3$Sn$_2$S$_2$, the blue sites correspond to Co atoms and the yellow sites to S atoms which hybridise with the kagome lattice of Co, driving the breathing distortion.}
\label{fig: hoppings}
\end{figure}

The kinetic model reduces to the ordinary nearest-neighbor kagome model in the case of $t_1 = t_2$, $t_3 = t_4 = 0$, which exhibits an ordinary \VHS{} at the $M$-point.
In comparing with DFT, a fitting choice reported in Ref. \cite{nag2024pomeranchuk} produces a set of hopping parameters $t^{Fit}_{1\rightarrow 4} \approx (-1.5, -0.5, 0.312, -0.375)\,t$. We normalize all kinetic parameters such that $t_1+t_2 = 2 t$. The average of $t_1$ and $t_2$ is used as energy scale, i.e., $t\approx 0.08\,\mathrm{eV}$. This captures the overall bandstructure and results in a \HOVHS{} at the $M$ points~\cite{nag2024pomeranchuk}. The bandstructure seen in DFT is in fact closer to the \HOVHS{} scenario than the fitting parameter model; therefore, following Nag \emph{et~al.}~\cite{nag2024pomeranchuk}, we use the mathematically exact \HOVHS{} parameters (with the chemical potential at Van Hove filling) for our theoretical model and reserve the DFT hopping parameters for a stability analysis (cf.~\cref{sec:stability_analysis}) of the diffuse nesting mechanism.
An exact \HOVHS{} is captured by the choice $t_{1\rightarrow 4} \approx (-1.383, -0.617, 0.352, -0.176)\,t$ with the overall scale $t\approx 0.08$\,eV~\cite{nag2024pomeranchuk}. 
These parameters lead to a vanishing second derivative of the energy dispersion along the $y$-direction at $M_2 = \tfrac{2\pi}{a\sqrt{3}}(0,1)$. In particular, it is possible to show that the second derivative vanishes along a continuous line through $t_3=-2t_4$ in the $t_3-t_4$ parameter space when keeping $(t_1, t_2)=\frac{2}{\frac{161}{41} + \frac{7}{4}} (-\tfrac{161}{41}, -\tfrac{7}{4})t$ fixed~\cite{nag2024pomeranchuk}. The energy dispersion at the Van Hove points is described by $\varepsilon(M_2 +\bm{k}) = \alpha k_x^2 - \beta k_y^4$ with positive constants $\alpha, \beta $. The resulting \HOVHS{} is classified in \cref{sec:kinetic_models_low_energy} as belonging to the $A_3$ class~\cite{Yuan2020classification, Chandrasekaran2020}.

\subsection{Non-breathing kagome model}
It is possible to achieve the same type of \HOVHS{} even without the breathing distortion in our model, i.e. with $t_1 = t_2$. One possible choice is $(t_1, t_2, t_3, t_4) = (-1, -1, 0.370845, -0.127329)\,t$. This restores the full $C_{6v}$ symmetry of the original kagome lattice. In analogy to the breathing scenario, a continuous line is present in the $t_3-t_4$ parameter space preserving the vanishing of the quadratic part.

\begin{figure}[]
\begin{center}
\includegraphics[width=0.75\columnwidth]{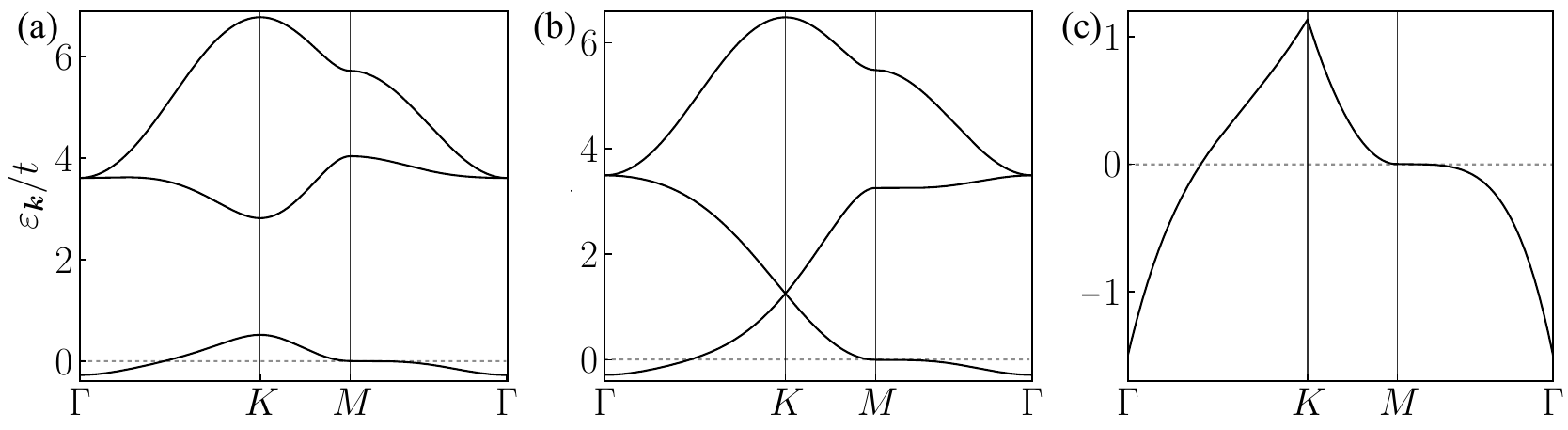}
\end{center}
\vspace{-0.3cm}
\caption[]{Bandstructures of (a) breathing kagome, (b) non-breathing kagome, and (c) Taylor expanded model of the lowest band. }
\label{fig:bandstructSM}
\end{figure}

\subsection{Low energy effective model}\label{sec:kinetic_models_low_energy}
To track the origin of diffuse nesting to the Van Hove singularities, we employ a Taylor approximated version of the above model. To this end, we project the model to its lowest band that contains the \HOVHS{}. We check that this is allowed by investigating the quantum metric of the lowest band, see \autoref{sec:QGT} for details. The resulting single-band energy dispersion is expanded around the M points up to 4\textsuperscript{th} order, which captures all leading contributions.  To find the canonical dispersion~\cite{Yuan2020classification}, i.e.\lk{,} the leading scale-invariant part of the dispersion defined by the quasi-homogeneity condition  $E(\lambda^a k_x, \lambda ^b k_y) = \lambda E(k_x, k_y)$, we identify the scaling exponents of the \HOVHS{} as $a=\frac{1}{2}, b=\frac{1}{4}$. This leads to a canonical dispersion of the form $E(k_x, k_y) = Ak_x^2+Bk_xk_y^2+Ck_y^4$, which corresponds to the $A_3$ class~\cite{Yuan2020classification, Chandrasekaran2020}. The prefactors of the \HOVHS{} in the breathing Kagome model are $(A, B, C) = (0.25871, 0, -0.00864)\,t$ when the axes are chosen such that the $k_x$ direction is on the $M K$ line and the $k_y$ direction on the $M\Gamma$ line.
To define the energy dispersion of the Taylor model on the complete hexagonal Brillouin zone, we divide the latter into six equilateral triangles. Each of them has a single $M$ point on its boundary. We define the energy dispersion as a piecewise function on these triangles, where in each triangle the canonical dispersion of the corresponding $M$ point is used. This approach conserves the lattice symmetry and ensures that the dispersion is continuous. It is, however, not differentiable on the $\Gamma K$ lines.

\section{Details of the FRG calculations}

\label{sec:FRG}
We employ the functional renormalization group in its standard static four-point approximation, neglecting self-energy corrections, frequency dependencies, and all interaction vertices with more than 4 fermionic fields, i.e. two interacting particles~\cite{Salmhofer2001, Metzner2012, Platt2013f, Profe2022, Beyer2022}. Since self-energy corrections enter only at higher loop order in the renormalization of the scattering vertex, they typically remain subleading and do not affect the qualitative tendencies toward phase transitions considered here \cite{Metzner2012, classen2020competing}.
The main methodological improvement compared to previous studies of HOVHS~\cite{classen2020competing, Han2023, castro2023emergence, YiMingWu_2023, hsu2021spin} is that within the aforementioned approximations our FRG analysis systematically accounts for all possible ordering tendencies and fluctuations representable as fermionic bilinears with arbitrary momentum transfers $\boldsymbol{q}$
in an unbiased fashion. While secondary momenta $\boldsymbol{k}$ and $\boldsymbol{k}'$ are expanded in real-space form factors that are truncated at some distance $\alpha$ (truncated unity FRG formulation), transfer momenta are sampled on equidistant momentum points within the Brillouin zone, where the modern implementation used in the present work~\cite{Profe2022, Beyer2022, Profe2024div1, Profe2024div2} allows for high resolutions up to $\approx 25000$ momentum mesh points within the BZ. 

The FRG interpolates between energy scales by progressively integrating out interaction processes using a renormalization group (RG) cutoff scale $\Lambda$, which is lowered from a high initial value to zero. This gradual flow captures the low-energy physics step by step, naturally probing the Fermi surface at varying levels of broadening. The onset of an ordered phase is signaled by a divergence of the effective two-particle interaction vertex at a critical scale $\Lambda_c$. Since all quantum fluctuations above the critical scale are incorporated, a mean-field decomposition of the resulting microscopic model given by the effective four-point vertex into superconducting, charge, and magnetic expectation values, can be performed to obtain the symmetry breaking order. 

Since the FRG is only valid directly at the phase transition---where the divergence is encountered---the mean-field decomposition is performed at the phase boundary where $\Delta\rightarrow0$. Consequently, we can linearize the self-consistent gap equation, recasting it as an eigenvalue problem for each $\boldsymbol{q}$ directly at the phase transition.
The maximum of eigenvalues at each transfer momentum in the particle-hole channel of our model, indicating the leading fluctuation, is displayed on an exemplary momentum mesh in Fig.~2a in the main text. The associated eigenstate of a divergent eigenvalue defines the structure of the leading order parameter.

\subsection{Numerical details}
The FRG calculations were performed with the TUFRG backend of the divERGe library \cite{Profe2024div1, Profe2024div2}, making use of the sharp frequency cutoff to boost numerical performance \cite{Profe2024div1}. We employed a $30\times30$ mesh for the bosonic momenta of the vertices, with an additional refinement of $51\times51$ for the integration of the loop. The form-factor cutoff distance is chosen as $3.01$ in units of lattice vector length. We check for convergence by calculating selected points in parameter space with an increased number of momentum points and form-factor cutoff (up to $150\times150$ with refinement $10\times10$, form factor cutoff up to $4.01$). The adaptive Euler integrator of the divERGe library is employed with default parameters.
\section{Phase diagrams and stability analysis}
\label{sec:phase_diagrams}
\subsection{Interaction parameter phase diagrams}
\begin{figure}[b!]
\begin{center}
\includegraphics[width=1 \columnwidth]{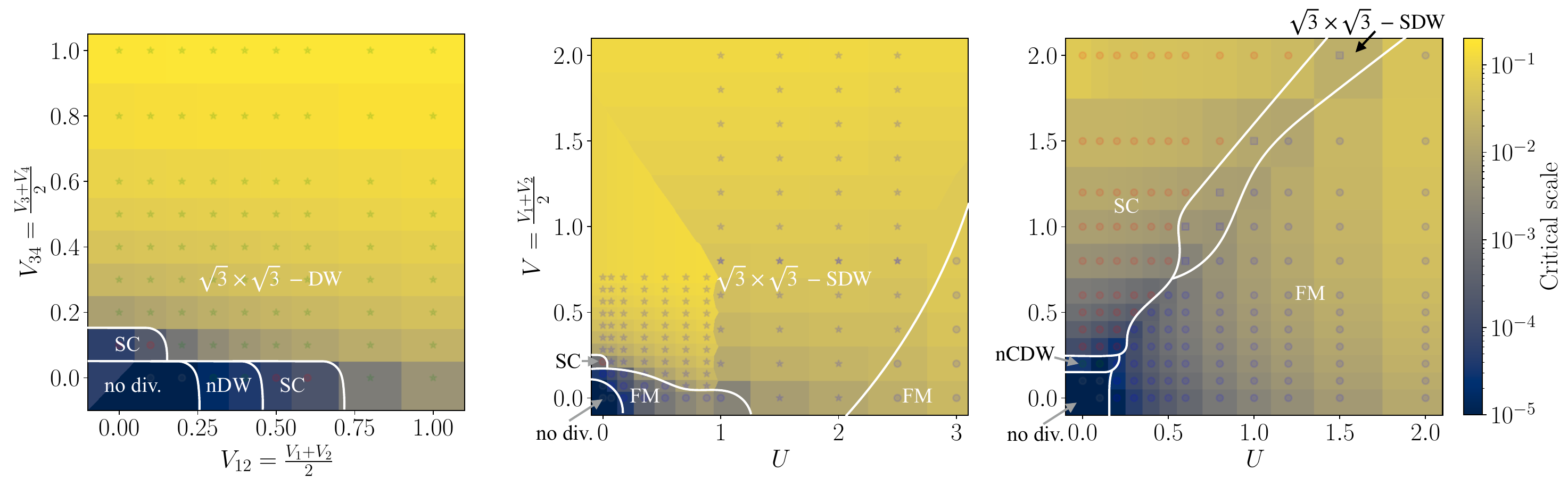}
\end{center}
\caption[]{Functional renormalization group interaction parameter space diagrams of the breathing configuration adapted from \textit{ab-initio} modeling of the Kagome surface termination of Co$_3$Sn$_2$S$_2$ \cite{nag2024pomeranchuk} with spinless (a) and spinful (b) fermions, as well as the spinful non-breathing model (c), introduced in \autoref{sec:kinetic_models}. While $\sqrt{3}\times\sqrt{3}$-order dominates across a broad parameter regime in the breathing configuration (a, b), its parameter window narrows in the non-breathing scenario.}
\label{fig:phase_diag}
\end{figure}

In general, identifying a suitable set of bare interaction parameters from first principles is challenging. While constrained random phase approximation (cRPA) calculations seem to suggest a universal scaling behavior for long-range interactions in several kagome compounds~\cite{disante2023electronic}, the present model features several key differences: The kagome bandstructure is constrained to the material's surface, is fully spin polarized, and exhibits a breathing distortion. Additionally, cRPA calculations are only suitable for predicting the relative strength of different interaction terms and do not allow for a faithful estimation of the absolute interaction scale~\cite{honerkamp2012effective, profe2025exact}.

To circumvent these problems, we perform FRG interaction parameter scans. We add density-density interactions to the tight-binding Hamiltonian up to fourth nearest neighbor distance:
\begin{equation}
        \mathcal{H}_{int} = U \sum_{i} n_{i, \uparrow} n_{i, \downarrow}+
        \sum_{n=1}^4 V_n \sum_{\langle i,j\rangle_n, \sigma \, \sigma'} n_{i, \sigma} n_{j, \sigma'}
\end{equation}
While for spinful fermions, we set $V_3 = V_4 = 0$ and only consider second nearest neighbor interactions, the absence of onsite $U$ in the spinless case naturally enhances the relevance of  long range interactions and necessitates the consideration of $V_{3,4}$~\cite{fu2024e, Zhan2024}.
In order to restrict the extensive interaction parameter space, we choose the ratios of interaction parameters $\frac{V_1}{V_2}$ and $\frac{V_3}{V_4}$ to resemble the ratios of the corresponding hybridization strengths. We checked the stability of the $K$ phase against variations of these ratios, including setting them to $1$. This leaves the relative strength of the pairs, $V_{12}$ and $V_{34}$, as free variables and reduces the interaction parameter space from four to two dimensions.

The resulting FRG phase diagrams of the breathing configuration adapted from \textit{ab-initio} modeling of the kagome surface termination of Co$_3$Sn$_2$S$_2$~\cite{nag2024pomeranchuk} with spinless (a) and spinful (b) fermions, as well as the spinful non-breathing model (c) introduced in \autoref{sec:kinetic_models}, are displayed in \autoref{fig:phase_diag}.

The predominant part of parameter space of the breathing model consists of the unprecedented $\sqrt{3}\times\sqrt{3}$ density wave order originating from the diffuse nesting mechanism introduced by the present work. In the limit of weak initial interactions, superconductivity and nematic orders emerge as the dominant tendencies. The latter is in agreement with previous theoretical studies of HOVHS scenarios in hexagonal lattice systems \cite{classen2020competing, Han2023} and theoretical and experimental studies of Co$_3$Sn$_2$S$_2$ \cite{nag2024pomeranchuk}. In experiment, such nematic order may be further stabilized by strain, deviations from the exact VH filling, or as a consequence of weak local correlations at the Co surface---situating the system in the weak coupling region of the phase diagram in which $\boldsymbol{q}=\Gamma$ orders dominate.
Upon introducing a spin degree of freedom, the leading instabilities reveals a magnetic ordering tendency rather than charge order; the alternating bond phase then corresponds to the spins rather than charge density. While we comprehensively analyzed the mean field equations and Ginzburg-Landau free energy for the scenario of spinless fermions, we leave a comparably thorough study of magnetic Kekulé order to future works.

When omitting the breathing configuration, therefore reintroducing sublattice polarization effects (see \autoref{sec:nesting}), we observe a suppression of $\sqrt{3}\times\sqrt{3}$-order \autoref{fig:phase_diag}(c), concurrently with a lowered critical scale.

\subsection{$K$-order stability analysis}\label{sec:stability_analysis}
To extend the stability analysis of the $K$-ordered phase beyond the interaction parameter regimes, we vary both the kinetic parameters and the chemical potential of the model given in \autoref{eqn:Hamiltonian_SM} and repeat the FRG analysis. The default parameters, used both in the theory section of Nag \emph{et~al.}~\cite{nag2024pomeranchuk} and our calculations, are those of the mathematically precise \HOVHS{}.
\subsubsection{Deviations from Van Hove filling} 
We observe a suppression of the $K$-ordering tendency upon small doping. While the order persists within the regime of intermediate initial interaction strengths for dopings $\Delta$$\mu \leq 0.05 t$, it is absent within the interaction parameter range accessible to the FRG calculation for $\Delta$$\mu \geq 0.1t$. A strong susceptibility to changes in ordering tendencies for small deviations from Van Hove filling is a well-known feature of FRG studies. Comparing the
suppression of instabilities with finite transfer momentum---induced by detuning the chemical potential---between the diffuse nesting scenario here with standard Van Hove
nesting, we find that in the diffuse nesting case a larger chemical potential change is tolerated.
The reference model with standard Van Hove nesting is a kagome Hubbard model at the $p$- and $m$-type Van Hove filling, analyzed via FRG calculations for various values of $\Delta \mu$ as performed in Ref.~\cite{Profe2024}.
The strong stability against detuning of the chemical potential, compared to other systems, is explained by its dependence on scattering events within a ``diffuse''{} energy window around the Fermi energy. Since this energy window is extended, a small deviation of chemical potential does not completely remove the overlap.
\subsubsection{Deviations from the higher-order Van Hove scenario}
As outlined earlier, a van Hove singularity is classified as higher-order if at least one of the Hessian eigenvalues vanishes. We analyze the effect of small deviations from this scenario by varying the hopping parameters of the model. To analyze the stability of the Kekulé order at the experimentally accessible DFT fit parameters that approximately correspond to Co$_3$Sn$_2$S$_2$~\cite{nag2024pomeranchuk} (see \autoref{sec:kinetic_models_breathing_kagome}), we interpolate between these and the exact \HOVHS{}. This reintroduces a finite quadratic term along the $\Gamma M$ line. At sufficiently flat quadratic dispersion, the $K$-order persists, demonstrating that the flatness-induced high DOS of the VHS close to its higher-order configuration drives the mechanism and an exact vanishing of the Hessian determinant is not required. As the deviation is increased, we observe that the transfer momentum associated with the ordering tendency shifts toward an incommensurate momentum point located along the $KM$ line. We find that this is consistent with a shift in the momentum of the particle-hole susceptibility peak, further substantiating the claim that diffuse nesting is purely of kinetic origin. However, if longer-range interactions that favor the charge imbalance between neighboring up-pointing triangles ($V_{34}$) are strong, the $K$-order can be stabilized at the DFT fit parameters even though the particle-hole susceptibility peak shifts to incommensurate momenta.
The stability of the $K$-order towards different hopping terms can be explained by the possibility to construct a universal single-band hexagonal $M$-HOVHS model, and by the diffuse nesting mechanism itself: Since we do not require an exact overlap of Fermi surfaces or of high-DOS points (as in perfect nesting or Van Hove nesting scenarios), the $K$-order is stable to a relatively large uncertainty in the hoppings.

\section{Origin of diffuse Fermi surface nesting }
\label{sec:nesting}
\subsection{Overlap of broadened Fermi surfaces}
\begin{figure*}[b]
\begin{center}
\includegraphics[width=0.65 \columnwidth]{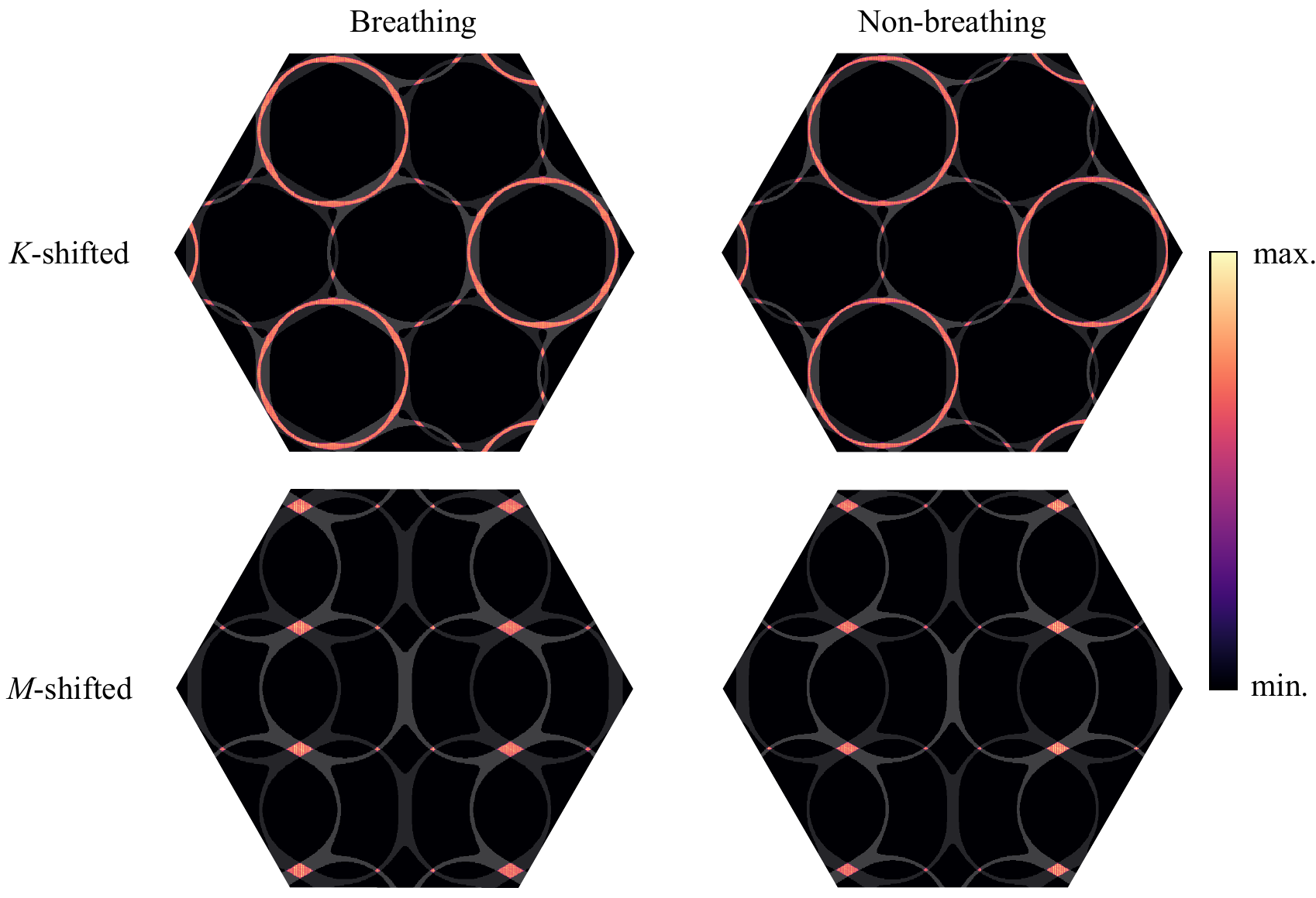}
\end{center}
\caption[]{Fermi surface overlaps of breathing and non-breathing kinetic models with nesting-vectors $\boldsymbol{q}= M$ and $\boldsymbol{q}=K$ at HOVHS filling. The broadening factor is $T=0.01\, t$. Upon $K$-shifting circular regions appear, indicating strong nesting. When shifted by $M$, the nesting is poor. 
}
\label{fig:Nesting_K_M_HOVH}
\end{figure*}

The particle-hole susceptibility peak at transfer-momentum $\boldsymbol{q}=K$ in Fig.~3a (main text) is traced back to the diffuse Fermi surface (FS) nesting inherent to the warped FS emerging in the presence of HOVHSs, where the quartic term points along the $\Gamma M$ direction. The Fermi surface overlaps at $\boldsymbol{q}=K$ and $\boldsymbol{q}=M$, with broadening realized by a finite temperature $T=0.01 t$, are shown as heatmaps in \autoref{fig:Nesting_K_M_HOVH}. The gray regions indicate the two shifted FS. In the $K$-shifted scenario extended circular nesting regions that drive a susceptibility peak appear, while the nesting at $M$-shifting remains poor. This is in contrast to the perfect VHS nesting emerging for ordinary VHS at transfer momenta $M$ in hexagonal lattice systems with only (real) nearest neighbor hybridization.

\subsection{Causes of Fermi surface broadening in kinetic and interacting models}
It is important to clearly distinguish the FS broadening in the kinetic models, where it is implemented by hand to investigate its effects on susceptibilities, from that in the interacting models of the FRG calculation.

In the kinetic model, the susceptibility in band space (cf.~Eq.~(4) in the main text)
% (cf.~\cref{eqn:bare_susc})
reduces, after evaluation of the Matsubara sum, to the Lindhard formula:
\begin{equation}
    \chi^0_{nm}(\bvec q) = - \frac{1}{\beta} \int_\mathrm{BZ} \text{d}\bvec k\, \frac{f(\epsilon_n(\bvec k + \bvec q)) - f(\epsilon_m(\bvec k))}{\epsilon_n(\bvec k + \bvec q) - \epsilon_m(\bvec k) + i\delta^+},
\end{equation}
where $f$ is the Fermi-Dirac distribution and $\epsilon_n(\bvec{k})$ denotes the dispersion of the $n$-th band.
There are multiple ways to broaden the effective energy window of the dominant scattering processes: The first is finite temperature, which modifies the numerator by ``smearing out''{} the Fermi functions. The other is a finite quasiparticle lifetime due to interactions or disorder that introduce an imaginary part to the denominator, modeled with a finite $\delta^+$, which results in a Lorentzian broadening. Both variations have the same effect of driving diffuse $K$-nesting by allowing scattering at non-zero energy.

In the FRG calculation for the interacting model, the RG scale $\Lambda$ effectively acts as broadening, playing a similar role as temperature~\cite{beyer2022reference, klebl2023competition, fischer2024theory}. Even though our approximations for the FRG fix $T=0$, this effectively broadens the Fermi surface and therefore allows for diffuse nesting.

Thermal broadening, which is used throughout the main section, is only one possibility to implement a broadened Fermi surface within the non-interacting models. While therefore a finite temperature is not expected to harm the diffuse nesting mechanism, it is not required, and the mechanism is expected to be stable at zero temperature.

\begin{figure*}[]
\begin{center}
\includegraphics[width=0.85 \columnwidth]{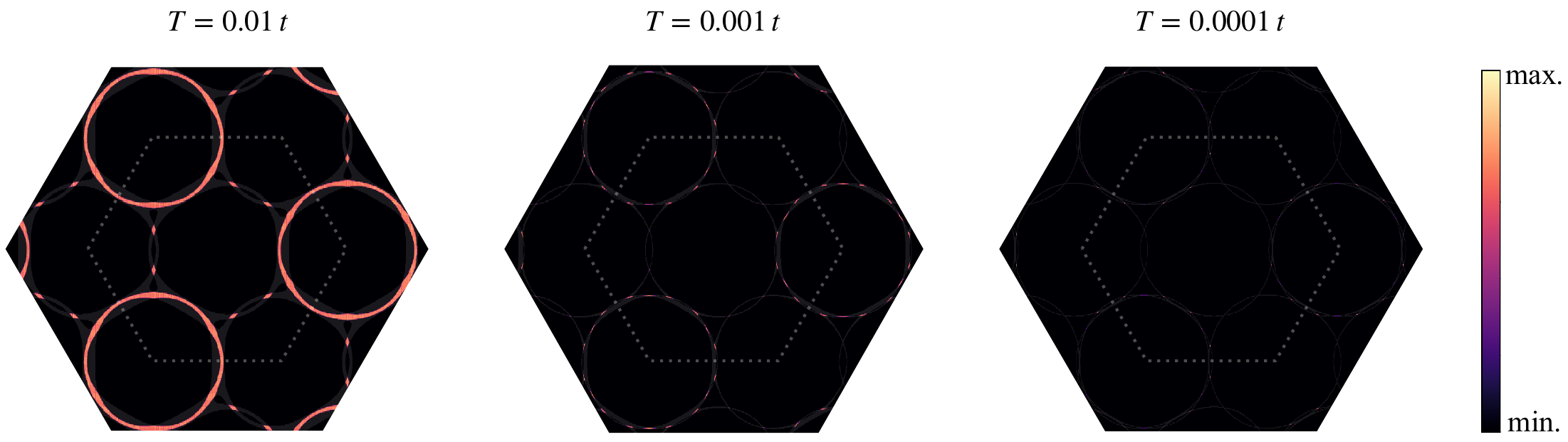}
\end{center}
\caption[]{Fermi surface nesting of the breathing configuration with nesting vector $\boldsymbol{q}=K$ at HOVHS filling with different broadening values $T$. When $T$ becomes infinitesimal the nesting condition vanishes.}
\label{fig:Nesting_K_HOVH}
\end{figure*}

Within these non-interacting models, upon lowering the FS broadening, the $\Gamma$-peak in the particle-hole susceptibilty surpasses the apex at $\boldsymbol{q}=K$. Visualizing the FS overlap of a $K$-shifted FS for different values of temperature broadening $T$ (\autoref{fig:Nesting_K_HOVH}) in the breathing configuration elucidates this observation. In the absence of broadening, i.e. a weak-coupling scenario, the overlaps vanish because of non-matching curvatures of the FS regions.

\section{Symmetry classification of $K$-orders on the kagome lattice}
\label{sec:symm}

Any order parameter can be categorized into irreducible representations (irreps) according to its transformation behavior under the symmetry operations of the crystal lattice.
The full symmetry group of the crystal lattice can be split into a set of discrete lattice translations $\mathcal{T}$ and pointgroup operations $\mathcal{P}$, such that the total symmetry group of the crystal can be written as $\mathcal{S} = \mathcal T \otimes \mathcal P$.
In the case of finite $Q$ orders, both the translational and pointgroup symmetry is partially reduced and the instability can be labeled by the order vector $Q$ and an irrep of its corresponding little group $\mathcal{P}_Q$. To be able to classify the resulting symmetry broken orders by a single irrep instead of a composite irrep of $\mathcal{T}$ and $\mathcal{P}$, the concept of extended pointgroups was introduced (cf. Ref.~\cite{Basko2008t, Venderbos2016s}): By enlarging the unit cell, $Q$ is mapped back to the zone center of the backfolded BZ and symmetry elements are moved from $\mathcal{T}$ to the extended pointgroup $\mathcal{P}^\prime$, which now also includes translations within the enlarged unit cell as generators of the group.
Thereby, the translation symmetry is not reduced at the phase transition, \textit{i.e.} $\mathcal T$ remains intact and the order parameter can be exclusively classified by irreps of $\mathcal{P}^\prime$. 

\begin{figure}[t]
\begin{center}
\includegraphics[width=0.5 \columnwidth]{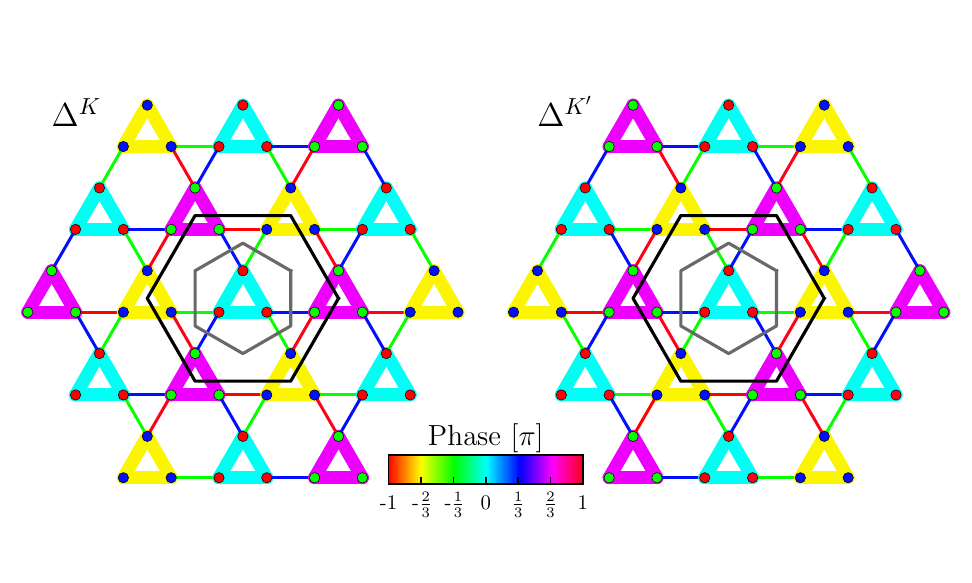}
\end{center}
\caption{Degenerate leading FRG eigenvectors in the Kekulé phase at $K$ and $\Kp$ up to nearest neighbor bond distances. The size (color) of the sites and bonds indicate the absolute value (complex phase) of the order parameter $\Delta^{K^{(\prime)}}_{o_1 o_2}(\bvec{r}) = \frac{1}{N_{\bvec k}} \sum_{\bvec{k}} e^{i \bvec{k} \cdot \bvec{r}} \, \Delta^{K^{(\prime)}}_{o_1 o_2}(\bvec{k})$. The original (extended) Wigner-Seitz cell is indicated in grey (black).}
\label{fig:order_parameter_K_complex}
\end{figure}

In the case of $K$-ordering, the appropriate symmetry group is no longer $C_{3v}$ of the original lattice, but the extended pointgroup $C_{3v}^{\prime \prime}$ of the enlarged $\sqrt{3} \times \sqrt{3}$ unit cell. 
When comparing the transformation behavior of the two states in \autoref{fig:order_parameter_K_complex} with the character table \autoref{tab:character_table}, the order parameter can be labeled by the $E_3^\prime$ irrep.

For a symmetry group with multiple equivalent rotation centers like $C_{3v}$ (equivalent to the wallpaper group $P3m1$), the angular momentum for the center of mass coordinate of the order becomes dependent on the choice of the rotation center.
E.g. if we choose the $C_3$ center in \autoref{fig:order_parameter_K_complex} as the center of a strong bond triangle, the obtained order acquires a trivial transformation behavior under rotation in contrast to the phase winding around the center of the hexagon.

\begin{table}[]
\centering
\caption{Character table of the extended pointgroup $C_{3v}^{\prime \prime}$ of the enlarged $\sqrt{3} \times \sqrt{3}$ unit cell of the breathing kagome lattice.
$E$ is the identity operation, $t_{a_{1, 2}}$ label translations with the primitive lattice vectors of the original kagome lattice, $C_3^{(2)}$ positive (negative) rotations by $2 \pi / 3$ and $\sigma_{1,2,3}$ are the three symmetry equivalent mirrors through the sides of each small triangle. We use the center of the hexagonal plaquettes as the symmetry center and note that a different rotation center leads to a permutation of the characters of the irreps $E_i^\prime$.}
\label{tab:character_table}
\vspace{0.5em}
\begin{ruledtabular}
\def\arraystretch{1.5}
\begin{tabular}{c|cccccc}
     $C_{3v}^{\prime \prime}$ &
     $E$ & $t_{a_1}, t_{a_2}$ & $C_3, C_3^2$ & $t_{a_1} C_3, t_{a_2} C_3^2$ & $t_{a_1} C_3^2, t_{a_2} C_3$ & $\sigma_{1,2,3}, t_{a_1} \sigma_{1,2,3}, t_{a_2} \sigma_{1,2,3}$ \\ \hline
     $A_1$ & 1 & 1 & 1 & 1 & 1 & 1 \\
     $A_2$ & 1 & 1 & 1 & 1 & 1 & -1 \\
     $E$ & 2 & 2 & -1 & -1 & -1 & 0 \\
     $E_1^\prime$ & 2 & -1 & 2 & -1 & -1 & 0 \\
     $E_2^\prime$ & 2 & -1 & -1 & 2 & -1 & 0 \\
     $E_3^\prime$ & 2 & -1 & -1 & -1 & 2 & 0
\end{tabular}
\end{ruledtabular}
\end{table}

\section{Free energy analysis of the resulting $K$-order}
\label{sec:freeE}

Since the transfer momenta $K$ and $\Kp$ are related by time reversal symmetry (TRS), the corresponding order parameters $\Delta^K$ and $\Delta^{\Kp}$ are degenerate eigenstates of the FRG vertex, which preserves all symmetries of the bare Hamiltonian throughout the flow, and obey
\begin{equation}
    \Bar{\Delta}^K_{o_1 o_2}(\bvec{k}) = \Delta^{\Kp}_{o_2 o_1}(\bvec k + K) \ .
\label{eqn:K_symmetry}
\end{equation}
Equivalently, both states are degenerate solutions of the linearized gap equation of the low-energy effective theory, which is valid directly at the transition point to the symmetry broken phase.
To determine the physically realized linear superposition inside the symmetry-broken state, we minimize the system's free energy within the mean field approximation.

From the FRG analysis, we know that the non-vanishing expectation value in the symmetry broken state is given by $\langle \cre{c}_{\bvec{k}+\bvec{q}} \ann{c}_k \rangle$ with $\bvec{q} \in \{K, \Kp \}$ and we can write down the corresponding mean field decoupling for the interacting Hamiltonian
\begin{equation}
\begin{split}
    H ={} & \sum_{\bvec{k} o_1 o_2} \mathcal{H}^0_{o_1 o_2}(\bvec{k}) \cre{c}_{\bvec{k} o_1} \ann{c}_{\bvec{k} o_2} +
    \frac{1}{N_{\bvec{k}}} \sum_{\bvec{q} \bvec{k} {\bvec{k}'} \{o_i\}} V^D_{o_1o_2o_3o_4}(\bvec q, \bvec{k}, {\bvec{k}'})
                                \cre{c}_{\bvec{k}+\bvec{q} o_1} \ann{c}_{\bvec{k} o_2} \cre{c}_{{\bvec{k}'}-\bvec{q} o_3} \ann{c}_{{\bvec{k}'} o_4} \\
\simeq{}& \begin{multlined}[t]
    \sum_{\bvec{k} o_1 o_2} \mathcal{H}^0_{o_1 o_2}(\bvec{k}) \cre{c}_{\bvec{k} o_1} \ann{c}_{\bvec{k} o_2} +
     \frac{1}{N_{\bvec{k}}} \sum_{\bvec{q} \in \{K, \Kp \}} \sum_{\bvec{k} {\bvec{k}'} \{o_i\}} V^D_{o_1o_2o_3o_4}(\bvec q, \bvec{k}, {\bvec{k}'})
        \Big[\langle \cre{c}_{\bvec{k}+\bvec{q} o_1} \ann{c}_{\bvec{k} o_2} \rangle \cre{c}_{{\bvec{k}'}-\bvec{q} o_3} \ann{c}_{{\bvec{k}'} o_4} + {}\\
        \cre{c}_{\bvec{k}+\bvec{q} o_1} \ann{c}_{\bvec{k} o_2} \langle \cre{c}_{{\bvec{k}'}-\bvec{q} o_3} \ann{c}_{{\bvec{k}'} o_4} \rangle    
        \Big] + C
    \end{multlined} \\
    = {}& \sum_{\bvec{k} o_1 o_2} \mathcal{H}^0_{o_1 o_2}(\bvec{k}) \cre{c}_{\bvec{k} o_1} \ann{c}_{\bvec{k} o_2} +
    2 \sum_{\bvec{q} \in \{K, \Kp \}} \sum_{\bvec{k} \{o_i\}} \Delta^{\bvec q}_{o_1 o_2}(\bvec{k}) \cre{c}_{\bvec{k}+\bvec{q} o_1} \ann{c}_{\bvec{k} o_2} + C = H_\mathrm{MF}\ ,
\label{eqn:mf_hamiltonian}
\end{split}
\end{equation}
where we have neglected second-order fluctuations around the MF order parameter
\begin{equation}
    \Delta^{\bvec q}_{o_1 o_2}(\bvec{k}) = \frac{1}{N_{\bvec{k}}} \sum_{{\bvec{k}'} o_3 o_4}
        V^D_{o_1 o_2 o_3 o_4}(\bvec q, \bvec{k}, {\bvec{k}'}) \langle \cre{c}_{{\bvec{k}'} - \bvec{q} o_3} \ann{c}_{{\bvec{k}'} o_4} \rangle \,.
\end{equation}
The quadratic term
\begin{equation}
\begin{split}
    C ={} & - \frac{1}{N_{\bvec{k}}} \sum_{\bvec{q} \in \{K, \Kp \}} \sum_{\bvec{k} {\bvec{k}'} \{o_i\}} V^D_{o_1o_2o_3o_4}(\bvec q, \bvec{k}, {\bvec{k}'})
    \langle \cre{c}_{\bvec{k}+\bvec{q} o_1} \ann{c}_{\bvec{k} o_2} \rangle \langle \cre{c}_{{\bvec{k}'}-\bvec{q} o_3} \ann{c}_{{\bvec{k}'} o_4} \rangle \\
    ={} & - \frac{1}{N_{\bvec{k}}} \sum_{\bvec{q} \in \{K, \Kp \}} \sum_{\bvec{k} {\bvec{k}'} \{o_i\}} \big(V^{D}_{o_1o_2o_3o_4}(\bvec q, \bvec{k}, {\bvec{k}'})\big)^{-1} \Delta_{o_1 o_2}^{-\bvec{q}}(\bvec{k}) \Delta_{o_3 o_4}^{\bvec{q}}({\bvec{k}'})    
\end{split}
\end{equation}
does not mix the different ordering fields and is hence constant for all states within the degenerate eigenspace of the FRG vertex with fixed total amplitude $|\Delta|^2 = |\Delta^K|^2 + |\Delta^{\Kp}|^2$.

Using the spinor notation $\cre{\psi}_{\bvec{k} o_1} = \big( \cre{c}_{\bvec{k} o_1}, \cre{c}_{\bvec{k}+K o_1}, \cre{c}_{\bvec{k}+\Kp o_1} \big)$, the MF Hamiltonian can be written as
\begin{equation}
    H_\mathrm{MF} = \frac{1}{3} \sum_{\bvec{k} o_1 o_2}
    \cre{\psi}_{\bvec{k} o_1}
    \begin{pmatrix}
        \mathcal{H}^0_{o_1 o_2}(\bvec{k}) & 2 \Delta_{o_1 o_2}^{\Kp}(\bvec{k} + K) & 2 \Delta_{o_1 o_2}^{K}(\bvec{k} + \Kp) \\
        2 \Delta_{o_1 o_2}^{K}(\bvec{k}) & \mathcal{H}^0_{o_1 o_2}(\bvec{k} + K) & 2 \Delta_{o_1 o_2}^{\Kp}(\bvec{k} + \Kp) \\
        2 \Delta_{o_1 o_2}^{\Kp}(\bvec{k}) & 2 \Delta_{o_1 o_2}^{K}(\bvec{k} + K) & \mathcal{H}^0_{o_1 o_2}(\bvec{k}+ \Kp)
    \end{pmatrix}
    \ann{\psi}_{\bvec{k} o_2} + C 
    = \frac{1}{3} \sum_{\bvec{k} n} \xi_n(\bvec{k}) \cre{\psi}_{\bvec{k} n} \ann{\psi}_{\bvec{k} n} + C \,,
\end{equation}
and we obtain the quasi-particle spectrum $\xi_n(\bvec{k})$ within the ordered phase by diagonalizing the Hamiltonian.
We note that the symmetry constraint of \autoref{eqn:K_symmetry} guarantees a hermitian MF Hamiltonian.
Exploiting the fermionic nature of the quasi-particles, i.e., $\cre{\psi}_{\bvec{k} n} \ann{\psi}_{\bvec{k} n} = \{ 0, 1 \}$, the free energy can by directly calculated as
\begin{equation}
    F = - T \ln(Z)
      = -T \ln \Big( \prod_{\bvec{k} n} \sum_{n_{\bvec kn}=0}^{1} e^{-\frac{ \xi_{n}(\bvec{k})}{3 T}} \Big) + C
      = -T \sum_{\bvec kn} \ln \Big( 1 + e^{-\frac{\xi_{n}(\bvec{k})}{3 T}} \Big) + C \,.
\label{eqn:free_energy}
\end{equation}
We evaluate the free energy for all possible solutions within the degenerate eigenspace of the leading FRG instability.
Thereby, we keep the functional form of the order parameter at $K$ and $\Kp$ as the eigenstates of the FRG vertex and vary the relative amplitude and phase, such that
\begin{equation}
\Delta_{o_1 o_2}(\bvec{k}) = e^{i \phi_K} A_K \Delta^K_{o_1 o_2}(\bvec{k}) + e^{i \phi_{\Kp}} A_{\Kp} \Delta^{\Kp}_{o_1 o_2}(\bvec{k}) \, .
\end{equation}
Since \autoref{eqn:K_symmetry} is not enforced by TRS directly, but is rather a direct consequence of hermiticity and $K$ and $\Kp$ being time reversal partners, this requirement persists in the symmetry broken phase.
The order parameter in the $K$-phase features a finite site-local, and hence momentum independent, component. Thereby \autoref{eqn:K_symmetry} directly implies $A_K = A_{\Kp}$ and $\phi_K = - \phi_{\Kp}$ and we need to optimize the free energy with respect to the simplified order parameter
\begin{equation}
    \Delta_{o_1 o_2}(\bvec{k}) = A \big[ e^{i \phi}  \Delta^K_{o_1 o_2}(\bvec{k}) + e^{-i \phi} \Delta^{\Kp}_{o_1 o_2}(\bvec{k}) \big] \, .
\label{eqn:order_parameter}
\end{equation}

A MF analysis after an FRG flow technically requires a regularization of the divergent part of the two particle vertex via an inverse Bethe-Salpeter resummation~\cite{Wang2014c,O2024c}.
With this approach, all degrees of freedom are restored and the correct symmetry breaking on the mean field level is ascribed to a modified bare interaction.
As apparent from \autoref{eqn:mf_hamiltonian}, the effective two-particle interaction only enters the term $C$, which merely determines the absolute amplitude of the order parameter as a function of temperature, but cannot distinguish between different linear combinations within the degenerate manifold.
Hence, we only need to evaluate the kinetic term in \autoref{eqn:free_energy} as a function of $\phi$ for different values of $T \lesssim \Tc$ and $A$.
This can be done for the original theory without any further restrictions~\cite{Platt2013f}.

\begin{figure}[tb]
\begin{center}
\includegraphics[width=0.45 \columnwidth]{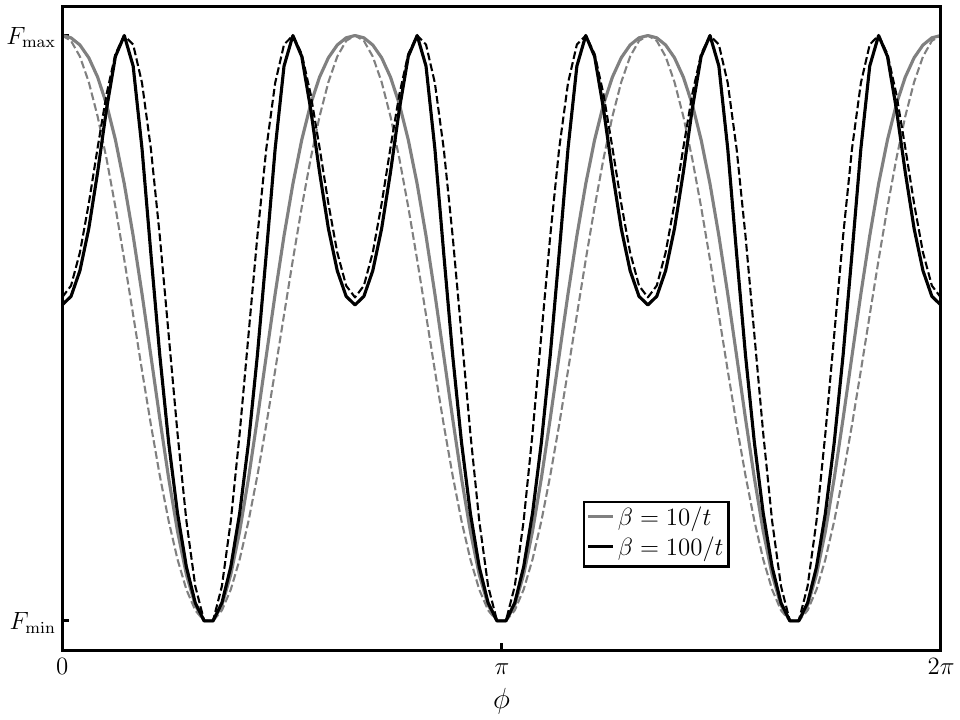}
\end{center}
\caption[]{Phase dependence of the free energy for different inverse temperatures $\beta = 1/T$ at fixed amplitude $A = t$ (solid lines). $F$ was calculated according to \autoref{eqn:free_energy} on a $1000 \times 1000$ momentum grid in the downfolded BZ. The dependence can be fitted by taking into account third and sixth order terms in the GL expansion in \autoref{eqn:GL_fit}(dashed lines).}
\label{fig:free_energy}
\end{figure}

The free energy in \autoref{fig:free_energy} shows three distinct minima at $\phi \in \{\pi/3, \pi, 5 \pi/3 \}$ for all values of $A$ and $T$, which are related by a translation by an original primitive translation. Hence, all minima can be mapped back to the same state by a redefinition of the unit cell. The obtained state is depicted in 
% \cref{fig:evals_k}b
Fig.~2b
of the main text.

\subsection{Complementary Ginzburg-Landau analysis}

This result can be easily understood by expanding the free energy in the order parameter field in terms of a Ginzburg-Landau (GL) analysis: Close to the phase transition, the free energy reads
\begin{equation}
    F = \alpha \Delta^K \Delta^\Kp
      + \gamma ( \Delta^K \Delta^K \Delta^K + \Delta^\Kp \Delta^\Kp \Delta^\Kp)
      + \mathcal{O}(\Delta^4) \ ,
      %+ \delta \Delta^K \Delta^\Kp \Delta^K \Delta^\Kp
      %+ \epsilon ( \Delta^K \Delta^K \Delta^K \Delta^K \Delta^\Kp + \Delta^\Kp \Delta^\Kp \Delta^\Kp \Delta^\Kp \Delta^K) + \mathcal{O}(\Delta^6) \ ,
\label{eqn:GL_free_energy}
\end{equation}
where the summation over the internal momentum structure of $\Delta^{K^{(\prime)}}(\bvec{k})$ has been carried out and incorporated in the GL coefficients $\alpha - \gamma$.
While $\alpha \propto (T - \Tc)$ determines the absolute size of the order parameter $A$ close to $\Tc$, the third order term $\propto \mathrm{Re}(\gamma) \cos(3 \phi) -  \mathrm{Im}(\gamma) \sin(3 \phi)$ is responsible for fixing the overall phase of the condensate~\cite{Bunney}.

For $\beta=1/T=10$, i.e. small size of the order parameter, where the expansion in \autoref{eqn:GL_free_energy} is still valid, the phase dependence of the free energy can be appropriately fitted by setting $\mathrm{Re}(\gamma) > 0$ and $|\mathrm{Im}(\gamma) / \mathrm{Re}(\gamma)| \ll 1$.
For lower temperature, where higher order terms of the free energy start to matter, interstitial local minima of the free energy develop, which can be attributed to the next term in the GL expansion that features a $\phi$ dependence
\begin{equation}
    F^{(6)} = \eta (\Delta^K \Delta^K \Delta^K \Delta^K \Delta^K \Delta^K
                    + \Delta^\Kp \Delta^\Kp \Delta^\Kp \Delta^\Kp \Delta^\Kp \Delta^\Kp) \propto \mathrm{Re}(\eta) \cos(6 \phi) - \mathrm{Im}(\eta) \sin(6 \phi) \ .
\end{equation}
Similarly assuming $|\mathrm{Im}(\eta) / \mathrm{Re}(\eta)| \ll 1$ for the expansion coefficients and $\mathrm{Re}(\eta) < 0$, also the free energy for $k_B T = 0.01 t$ can be fitted satisfactorily using
\begin{equation}
    F_{GL} = A_3 \cos(3 \phi) + A_6 \cos(6 \phi) \ ,
\label{eqn:GL_fit}
\end{equation}
where $A_6/A_3 = 0 \, (-1.25)$ for $\beta = 10 \, (100)$.

\subsection{Quasiparticle bandstructures}

\begin{figure}[t]
\begin{center}
\includegraphics[width=0.45 \columnwidth]{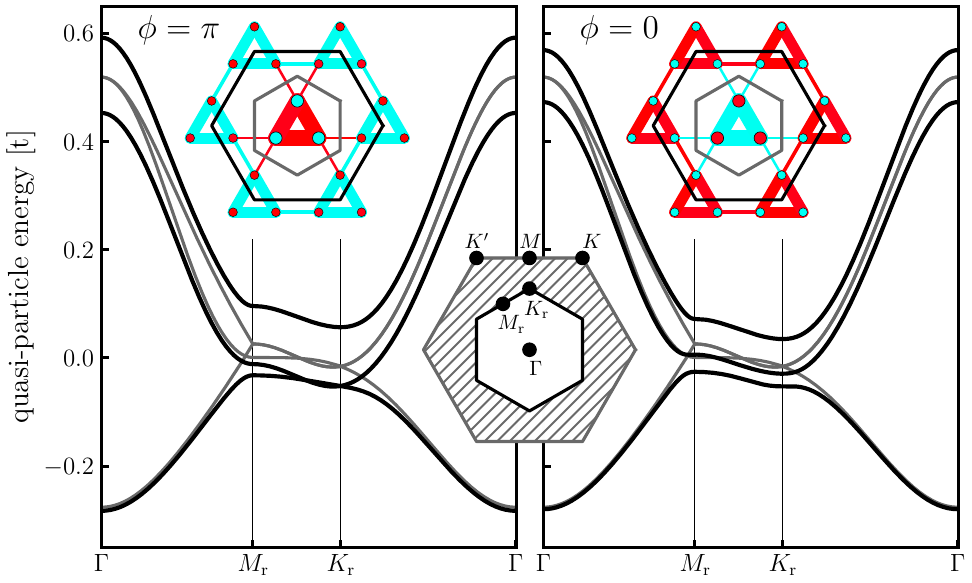}
\end{center}
\caption[]{Quasiparticle bandstructure of the breathing kagome for MF states with different global phases $\phi$. Grey lines indicate the bandstructure of the breathing kagome model backfolded from the original BZ (grey) to the reduced BZ (black) as shown in the inset. Black lines show the quasiparticle bandstructure for the order parameter in \autoref{eqn:order_parameter} with $A = 0.1 t$ and $\phi = \pi (0)$ in the left (right) panel. The insets display the order parameter in real space according to the convention in \autoref{fig:order_parameter_K_complex} with the original (enlarged) unit cell indicated in grey (black). Red (cyan) corresponds to negative (positive) values of the order parameter on the respective sites and bonds.}
\label{fig:qp_bandstructure}
\end{figure}

From the GL analysis, the importance of the global phase of the order parameter becomes directly apparent.
In the GL functional, this is reflected by the non-vanishing third order terms due to the relation $3 K = \Gamma$.
For $\phi_1 \in \{1/3 \pi, \pi, 5/3 \pi \}$ we have two stronger bonded and one weaker bonded triangle in the $\sqrt{3} \times \sqrt{3}$ unit cell, while for $\phi_2 \in \{0, 2 \pi/3, 4 \pi/3 \}$ the situation is reversed (cf. insets of \autoref{fig:qp_bandstructure}).
The structure of the leading FRG instability locks the $\pi$ phase shift between the onsite and bond components (cf. \autoref{fig:order_parameter_K_complex}).
This has direct consequences for the quasiparticle bandstructures obtained by diagonalizing the MF Hamiltonian in \autoref{eqn:mf_hamiltonian}, where the quasiparticle gap at the $\Gamma$ point (formerly $K^{(\prime)}$) is significantly enhanced for $\phi_1$.
The Kekulé charge order destroys the higher order nature of the vHs with a flat dispersion along the $M=M_r-K_r$ direction and induces a transition to an ordinary vHs. As a result, significant spectral weight is shifted away from the Fermi level, and the system lowers its free energy considerably. 

\section{Similarities to graphene-Kekul\'e order}
\label{kekule}

Our $\sqrt{3}\times\sqrt{3}$ order exhibits several defining features of Kekulé order, such as translation symmetry breaking, tripling of the unit cell, gapping of the Dirac point, momentum-space structure, and bond-order modulation, which go beyond a generic $\sqrt{3}\times\sqrt{3}$ order.

While a protecting symmetry to either site or bond ordering is absent in our model, and any interpolation between site and bond contribution can be found in a given charge or spin density wave susceptibility channel, we find the bond ordering part to largely dominante within our $\sqrt{3}\times\sqrt{3}$ order.

\begin{figure}[t!]
\begin{center}
\includegraphics[width=0.3 \columnwidth]{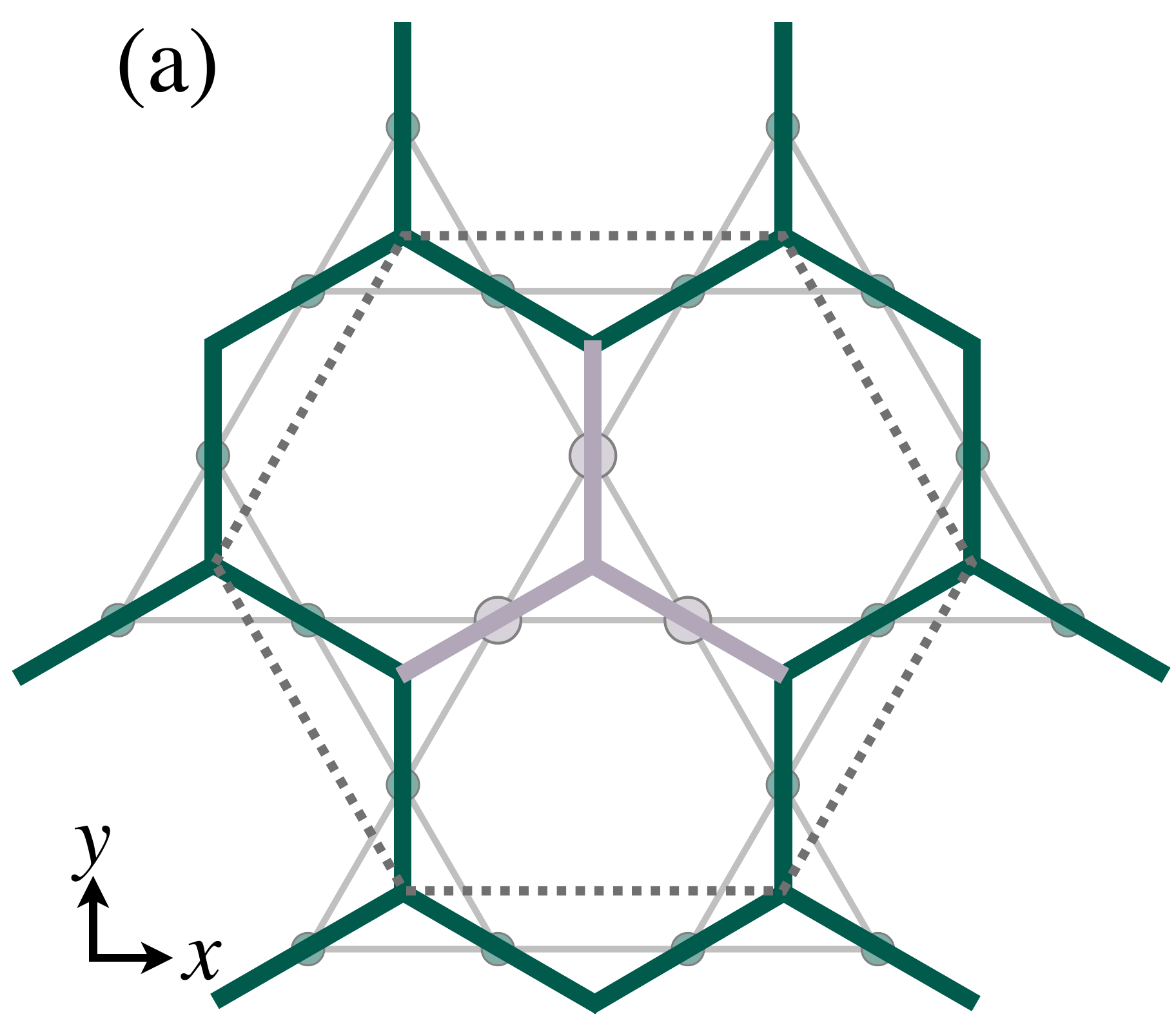}
\end{center}
\caption{Line graph reconstruction of the kagome lattice revealing a honeycomb structure. The bonds are colored corresponding to the onsite values of the associated onsite kagome Kekul\'e order forming a Kekul\'e-Y order on the effective honeycomb lattice. }
\label{fig:Kekule}
\end{figure}

Since the concept of Kekul\'e order is etymologically based on the properties of the honeycomb lattice, it is curious to draw attention to the real space similarities between our charge ordered kagome Kekul\'e phase and those previously studied in honeycomb models. The kagome lattice can be constructed as the line graph of a honeycomb lattice. Inspecting the line graph reconstruction of the onsite part of the obtained $K$-order reveals a Kekul\'e-Y order \cite{Ganesh2024} on the effective honeycomb structure, as shown in \autoref{fig:Kekule}. From these shared properties, we conclude that the observed order can be rigorously identified as Kekul\'e order.

\section{Quantum geometry}
\label{sec:QGT}
\subsection{Quantum metric}
The quantum metric $g_{\mu\nu}^{(n)}(\boldsymbol{k})$ for band $n$ is defined as the real part of the quantum geometric tensor $Q_{\mu\nu}^{(n)}(\bvec{k}) =
    \langle \partial_{k_\mu} u_n(\bvec{k}) |
    \left( 1 - |u_n(\bvec{k})\rangle \langle u_n(\bvec{k})| \right)
    | \partial_{k_\nu} u_n(\bvec{k}) \rangle$ with the Bloch function of the $n$-th band $|u_n(\bvec{k})\rangle$. It defines the infinitesimal distance between nearby states in Hilbert space as $ds_n^2(\bvec{k}) =
    g_{\mu\nu}^{(n)}(\bvec{k})\, dk_\mu\, dk_\nu$. Its Brillouin zone integral
\begin{equation}
    \Omega_{\mathrm{I}} =\int_{\mathrm{BZ}} \frac{d^2\boldsymbol{k}}{V_{\mathrm{BZ}}} \, \mathrm{Tr}\, [g(\boldsymbol{k})]
\end{equation}
provides a lower bound for the spread of Wannier functions~\cite{yu2025quantumgeometryquantummaterials, witt2025quantumgeometrylocalmoment}. For our purposes, this is important because an isolated band with a small value of $ \Omega_{\mathrm{I}}$ admits a description in terms of well-localized Wannier functions. Consequently, the system can be accurately reduced to a single-orbital tight-binding model for that band, which we require to construct a meaningful low-energy effective model (\autoref{sec:kinetic_models_low_energy}).

We analyze $\Omega_{\mathrm{I}}$ for different magnitudes of breathing, corresponding to varying gap sizes between the band hosting the HOVHS and the two residual bands. A linear interpolation of the kinetic parameters is employed between the non-breathing HOVHS kagome lattice [\autoref{fig:bandstructSM}(b)] and the breathing case [\autoref{fig:bandstructSM}(a)].

\begin{figure}[b!]
\begin{center}
\includegraphics[width=0.4 \columnwidth]{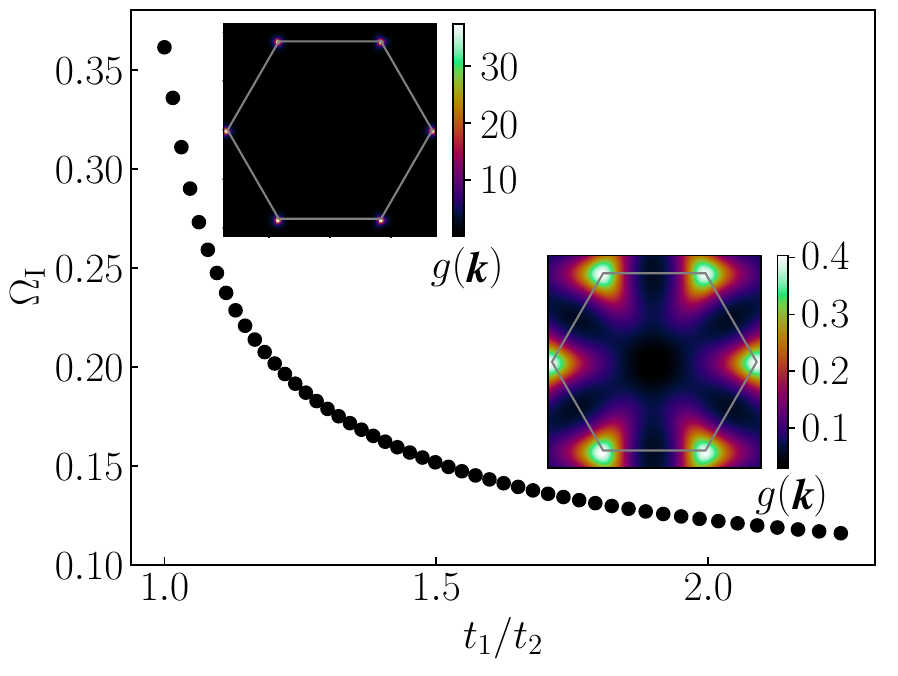}
\end{center}
\caption{Quantum metric (insets) and Brillouin zone integral over the quantum metric for linear interpolation between non-breathing \autoref{fig:bandstructSM}(b) and breathing \autoref{fig:bandstructSM}(a) kagome lattice.}
\label{fig:QGT}
\end{figure}
\autoref{fig:QGT} displays the evolution of $\Omega_{\mathrm{I}}$ from an almost non-breathing configuration ($t_1/t_2 \approx 1$) to the breathing model discussed in the main text. The decrease in $\Omega_{\mathrm{I}}$ indicates a stronger localization of the Wannier functions. The insets show the quantum metric in the BZ for $t_1/t_2 \approx 1.06$ and $t_1/t_2 \approx 2.24$. While being highly localized in momentum space for the nearly non-breathing case, the quantum metric becomes lower in magnitude and more spread out across the BZ in the breathing scenario, corroborating the behavior observed in $\Omega_{\mathrm{I}}$.

\subsection{Effects of sublattice polarization}

The lattice geometry of the unmodified kagome model leads to a complete (partial) sublattice polarization at the pure $p$- (mixed $m$-) type VHS at the $M$ points, protected by $\sigma_v$ mirror symmetry. Upon opening an infinitesimal gap, which is required to define a lower band and thus a quantum metric for it, the lower bound $\Omega_{\mathrm{I}}$ of the Wannier function spread is evaluated to be large. This indicates that a single-orbital description of the model is not possible.

In the breathing configuration, the $\sigma_v$ symmetry is explicitly broken on the single-particle level and the FS states feature support on all three sublattices throughout the Fermi surface. Hence, the nesting analysis in \autoref{fig:Nesting_K_M_HOVH} does not require a detailed examination of the electronic eigenstates on the FS. As the breathing introduces a trivial gap in the spectrum which is accompanied by decreased $\Omega_{\mathrm{I}}$ (see \autoref{fig:QGT}), it permits a single-orbital description of the lowest band hosting the HOVHS. This naturally excludes any sublattice polarization effects from emerging in the low-energy description of the system.

\begin{figure*}[t]
\begin{center}
\includegraphics[width=0.9 \columnwidth]{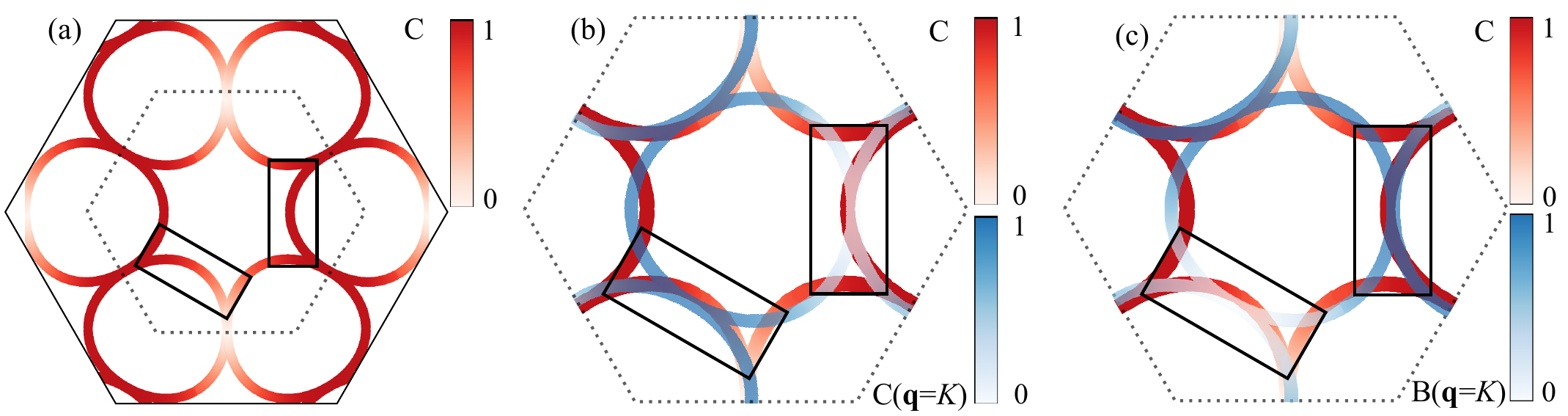}
\end{center}
\caption[]{Fermi surface nesting in the non-breathing scenario.
(a)~C sublattice weight along the Fermi surface, indicated by color. The sublattice polarization varies smoothly along the FS and is hence present over an extended region around the $M$-point, where a vanishing C weight is enforced by $C_2$ symmetry. 
 (b)~Original (red) and $K$-shifted (blue) Fermi surface. The opacity indicates again contribution of sublattice $C$ to the FS states. The FS regions with most weight on C only nests with FS segments with little C contribution (indicated by the black rectangle) reproducing the well-known sublattice interference effect also for $K$-nesting.
 (c)~Same as (b) but now with different sublattice contribution on the shifted Fermi surface, indicating the available scattering processes for nearest-neighbor interactions. Now FS states with high occupation on both the original and shifted FS nest to drive a $K$-ordering. 
}
\label{fig:sublattice}
\end{figure*}

In the fully $C_{6v}$ symmetric non-breathing \HOVHS{} scenario, the HOVHS is of mixed sublattice type and the FS around the three inequivalent $M$-points becomes depopulated of weight from a specific sublattice (cf. \autoref{fig:sublattice}(a)). As in the unmodified kagome model, this is accompanied by an increased lower bound of the Wannier function spread (see \autoref{fig:QGT}).
It is worth noting that due to the continuous evolution of the FS states, especially in the FS segments with large DOS close to the \HOVHS{} points along the $\sigma_v$ mirror plane, sublattice polarization is strong even away from the singularity.
This necessitates a sublattice-resolved analysis of the $\boldsymbol{q}=K$ nesting that differs substantially between the identical FS shapes in \autoref{fig:Nesting_K_M_HOVH}:
The overlap region highlighted in \autoref{fig:sublattice}(b) indicates a dominant FS segment of low sublattice weight, which is nested with regions of high weight of the same sublattice.
These states are therefore inaccessible to local scattering events and require a mediation by a NN repulsion that mixes sublattices (compare \autoref{fig:sublattice}(b) and (c)).
This scenario for the $K$-nesting is exactly analogous to the well-known sublattice interference effect for the $M$ point VHS nesting in the kagome Hubbard model~\cite{Kiesel2012a}. Consequently, we inspect a suppression of critical scales and a disappearing $K$-phase in the small $V_{12}$ regime in \autoref{fig:phase_diag}(c). For large NN repulsion, we recover the usual SC order known from FRG studies on the conventional kagome Hubbard model~\cite{Wang2013, Kiesel2013, Profe2024}.
While sublattice interference, as originally proposed, favors bond order formation by enhancing longer-range correlations, the simultaneous suppression of the critical scale disfavors the Kekul\'e bond order, which we have shown to arise from diffuse nesting.

\end{document}